\documentclass[twocolumn,showpacs,amsmath,amssymb]{revtex4}

\usepackage{graphicx}
\usepackage{dcolumn}
\usepackage{bm}

\newcommand{\brar}[1]{\left(#1\right)}

\newcommand{\dd}[1]{\frac{{\rm d}}{{\rm d}#1}}
\newcommand{\ddt}{\dd{t}}

\newcommand{\com}[2]{\left[#1,#2\right]}

\newcommand{\beq}{\begin{eqnarray}}
\newcommand{\eeq}{\end{eqnarray}}

\newcommand{\barr}{\begin{array}}
\newcommand{\earr}{\end{array}}

\newcommand{\half}{\frac{1}{2}}

\newcommand{\bra}[1]{\langle#1|}
\newcommand{\ket}[1]{|#1\rangle}

\newcommand{\op}[2]{\ket{#1}\bra{#2}}

\newcommand{\damping}[3]{\left(2#1#2#3-#3#1#2-#2#3#1\right)}

\newcommand{\phik}{\ket{\phi}}

\newcommand{\mean}[1]{\left\langle #1 \right\rangle}

\begin{document}

\title{Entangled-state cycles from conditional quantum evolution}

\author{Mile Gu}
 \altaffiliation{Present address: Department of Physics, University of Queensland, QLD 4072,
 Australia}
 \author{Scott Parkins}%
\author{H.~J.~Carmichael}%
\affiliation{%
Department of Physics, University of Auckland, Private Bag 92019, Auckland, New Zealand
}%

\date{\today}

\begin{abstract}
A system of cascaded qubits interacting via the oneway exchange of photons is studied. While for general
operating conditions the system evolves to a superposition of Bell states (a dark state) in the long-time
limit, under a particular \textit{resonance} condition no steady state is reached within a finite time.
We analyze the conditional quantum evolution (quantum trajectories) to characterize the asymptotic behavior
under this resonance condition. A distinct bimodality is observed: for perfect qubit coupling, the system
either evolves to a maximally entangled Bell state without emitting photons (the dark state), or executes
a sustained entangled-state cycle---random switching between a pair of Bell states while emitting a
continuous photon stream; for imperfect coupling, two entangled-state cycles coexist, between which a
random selection is made from one quantum trajectory to another.
\end{abstract}


\maketitle

\section{\label{sec:level1}Introduction}

Quantum entanglement is a feature of quantum mechanics that has captured much recent interest due to its
essential role in quantum information processing \cite{Nielson}. It may be characterized and manipulated
independently of its physical realization, and it obeys a set of conservation laws; as such, it is
regarded and treated much like a physical resource. 

It proves useful in making quantitative predictions to quantify entanglement.When one has complete
information about a bipartite system---subsystems $A$ and $B$---the state of the system is pure and there
exists a well established measure of entanglement---the \textit{entropy of entanglement}, evaluated as
the von Neumann entropy of the reduced density matrix,
\begin{equation}
E(\ket{\phi}_{AB}) = \textrm{Tr}(\rho_A \log_2 \rho_A),
\label{eq:von-Neumann}
\end{equation}
with $\rho_A\equiv\textrm{Tr}_B(\ket{\phi}\bra{\phi}_{AB})$. This measure is unity for the Bell states
and is conserved under local operations and classical communication. Unfortunately, however, quantum systems
in nature interact with their environment;  states of practical concern are therefore mixed, in which case
the quantification of entanglement becomes less clear.

Given an ensemble of pure states, $\{\ket{\phi_i}_{AB}\}$ with probabilities $\{ p_i\}$, a natural
generalization of $E(\ket{\phi}_{AB})$ is its weighted average $\sum_i p_i E(\ket{\phi_i}_{AB})$.
A difficulty arises, though, when one considers that a given density operator may be decomposed in 
infinitely many ways, leading to infinitely many values for this average entanglement. The density operator
for an equal mixture of Bell states $\ket{\Phi^\pm}=(\ket{0}_A\ket{0}_B\pm\ket{1}_A\ket{1}_B)/\sqrt{2}$,
for example, is identical to that for a mixture of $\ket{0}_A\ket{0}_B$ and $\ket{1}_A\ket{1}_B$, yet by the
above measure the two decompositions have entanglement one and zero, respectively.

Various measures have been proposed to circumvent this problem, most of which evaluate a lower bound. One
such measure, the \textit{entanglement of formation}, $E_F(\rho)$ \cite{bennet1}, is defined as the minimal
amount of entanglement required to form the density operator $\rho$, while the \textit{entanglement of
distillation}, $E_D(\rho)$ \cite{bennet2}, is the guaranteed amount of entanglement that can be extracted
from $\rho$.  These measures satisfy the requirements for a physical entanglement measure set out by
Horodecki \textit{et al}. \cite{horodecki3}. They give the value zero for $\rho_{AB}=(\ket{\Phi^+}
\bra{\Phi^+}+\ket{\Phi^-}\bra{\Phi^-})/2$, which might be thought somewhat counterintuitive, since this
state can be viewed as representing a sequence of random ``choices'' between two Bell states, both of which
are maximally entangled. This is unavoidable, however, because assigning $\rho_{AB}$ a non-zero value of
entanglement would imply that entanglement can be generated by local operations. The problem is fundamental,
steming from the inherent uncertainty surrounding a mixed state: the state provides an incomplete description
of the physical system, and in view of the lack of knowledge a definitive measure of entanglement cannot
be given.

An interacting system and environment inevitably become entangled. The problem of bipartite entanglement
for an open system is therefore one of tripartite entanglement for the system and environment. Complicating
the situation, the state of the environment is complex and unknown. Conventionally, the partial trace with
respect to the environment is taken, yielding a mixed state for the bipartite system. If one wishes for a
more complete characterization of the entanglement than provided by the above measures, somehow the inherent
uncertainty of the mixed state description must be removed.

To this end, Nha and Carmichael \cite{Nha04a} recently introduced a measure of entanglement for open systems
based upon quantum trajectory unravelings of the open system dynamics \cite{Carmichael93}. Central to their
approach is a consideration of the way in which information about the system is read, by making measurements,
from the environment. The evolution of the system conditioned on the measurement record is followed, and the
entanglement measure is then contextual---dependent upon the kind of measurements made. Suppose, for example,
that at some time $t$ the system and environment are in the entangled state
\begin{equation}
\ket{\phi} = \sum_{i,j}c_{i,j}\ket{\phi_i}_S\ket{\phi_j}_E.
\end{equation}
A partial trace with respect to $E$ yields a mixed state for $S$. If, on the other hand, an observer makes
a measurement on the environment with respect to the basis $\{\ket{\phi_j}_E\}$, obtaining the ``result''
$\ket{\phi_k}_E$, the reduced state of the system and environment is
\begin{subequations}
\begin{equation}
\ket{\phi^\prime}=\ket{\phi^\prime}_S\ket{\phi_k}_E,
\end{equation}
with conditional system state
\begin{equation}
\ket{\phi^\prime}_S=\sum_i c_{i,k}\ket{\phi_i}_S/\sqrt{p_k},
\end{equation}
\end{subequations}
where $p_k=\sum_i |c_{i,k}|^2$ is the probability of the particular measurement result. Thus, the system and
environment are disentangled, so the system state is pure and its bipartite entanglement is defined
by the von Neumann entropy, Eq.~(\ref{eq:von-Neumann}). Nha and Carmichael \cite{Nha04a} apply this idea to
the continuous measurement limit, where $\ket{\phi^\prime}_S$ executes a conditional evolution over time.

In this paper we follow the lead of Nha and Carmichael, also Carvalho \textit{et al.} \cite{Carvalho05},
not to compute their entanglement measure \textit{per se}, but to examine the entanglement \textit{dynamics} of
a cascaded qubit system coupled through the oneway exchange of photons. The system considered has been shown to
produce unconditional entangled states---generally a superposition of Bell states---as the steady-state solution
to a master equation \cite{Clark03}. For a special choice of parameters (resonance), a maximally entangled Bell
state is achieved $\ldots$ except that the approach to the steady state takes place over an infinite amount of time.

Here we analyze the conditional evolution of the qubit system to illuminate the dynamical creation of
entanglement in the general case, and to explain, in particular, the infinitely slow approach to steady-state
in the special case. We demonstrate that in the special case the conditional dynamics exhibit a distinct bimodality,
where the approach to the Bell state is only one of two possibilities for the asymptotic evolution: the second
we call an \textit{entangled-state cycle}, where the qubits execute a sustained stochastic switching between two
Bell states. Though involving just two qubits and elementary quantum transitions, the situation is similar to that
of a bimodal system in classical statistical physics in the limit of a vanishing transition rate between attractors.

The physical model of the cascaded qubit system is presented in Sec.~\ref{subsec2_1} and the quantum trajectory
unraveling of its conditional dynamics in Sec.~\ref{subsec3_1}. In Sec.~\ref{subsec4_1} we analyze the quantum
trajectory equations to demonstrate bimodality and the existence of entangled-state cycles. Finally, a discussion
and conclusions are presented in Sec.~\ref{subsec5_1}.

\section{\label{subsec2_1}The Cascaded Qubit System}

In this section we briefly outline the physical model for the cascaded qubit system to be analyzed. A more
detailed description, together with the techniques and assumptions used to derive the model master equation
presented here, is available in \cite{Clark03}.

\subsection{Physical Configuration}

\begin{figure}[b]
\begin{center}
\includegraphics[width=0.4\textwidth]{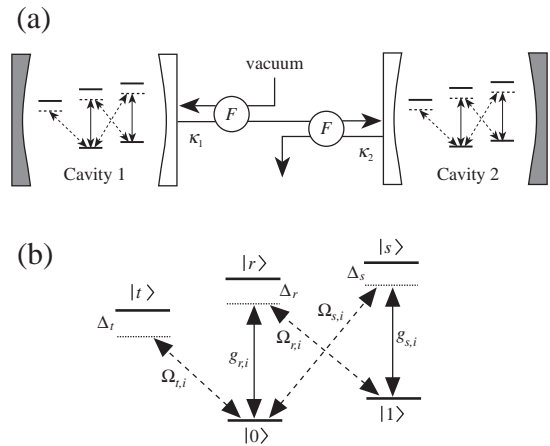}
\caption{(a) A pair of cascaded cavities, Cavity 1 and Cavity 2, each contain a single trapped atom;
a unidirectional coupling between the cavities is realized by Faraday isolators {\em F}. (b) The atomic
excitation scheme couples two stable ground states, $\ket{0}$ and $\ket{1}$, to three excited states,
$\ket{r}$, $\ket{s}$, and $\ket{t}$.}
\label{fig:fig1}
\end{center}
\end{figure}

The system considered consists of two high-finesse optical cavities, each containing a single tightly-confined
atom, the cavities arranged in a cascaded configuration with unidirectional coupling from cavity 1 to cavity 2
(Fig.~\ref{fig:fig1}). For simplicity, we consider the cavity modes to be identical, with resonance frequency
$\omega_{\rm cav}$ and field decay rate $\kappa$. Inefficiencies and losses in the coupling between the
cavities are modeled by a real parameter $\epsilon$, $0\le\epsilon\le1$, with perfect coupling corresponding
to $\epsilon=1$. The atoms are assumed to have five relevant electronic levels, of which two ground states,
$\ket{0}$ and $\ket{1}$, represent an effective two-state system, or qubit.

For each atom, the cavity field in combination with auxiliary laser fields (incident from the side of the
cavity) drives two separate resonant Raman transitions between states $\ket{0}$ and $\ket{1}$. An additional
laser field coupled to the $\ket{0}\leftrightarrow\ket{t}$ transition provides a tunable light shift of the
energy of state $\ket{0}$. All fields are assumed far detuned from the atomic excited states, so these states
may be adiabatically eliminated and atomic spontaneous emission ignored. Under the further assumption that
the cavity field decay rate is much  larger than the transition rates between $\ket{0}$ and $\ket{1}$, the
cavity fields may also be adiabatically eliminated to yield a master equation for the reduced two-atom
density matrix~$\rho$,
\begin{eqnarray}
\label{eq:me}
\dot\rho=\mathcal{L}\rho &=& \sum_{i=1,2}\damping{\hat R_i}{\rho}{\hat R_i^\dag}\nonumber\\*
&&-2\sqrt{\epsilon}\brar{\com{\hat R_1\rho}{\hat R_2^\dag}+\com{\hat R_2}{\rho\hat R_1^\dag}},
\end{eqnarray}
with
\begin{equation}
\hat R_i=(\beta_{r,i}\hat\sigma_{i-}+\beta_{s,i}\hat\sigma_{i+})/\sqrt{\kappa},
\end{equation}
where $\sigma_{i-}\equiv(\op{0}{1})_i$, and $|\beta_{r,i}|^2/\kappa$ and $|\beta_{s,i}|^2
/\kappa$ are the rates of $\ket{1}_i\rightarrow\ket{0}_i$ and $\ket{0}_i\rightarrow\ket{1}_i$
transitions, respectively.

By virtue of the cavity output, the system is an open system and solutions to master equation (\ref{eq:me})
generally describe mixed states. Under appropriate conditions, however, the system evolves to a pure and
entangled steady state.

\subsection{\label{subsec1_2}Steady State}

\begin{figure}[b]
\includegraphics[angle=0,width=0.4\textwidth]{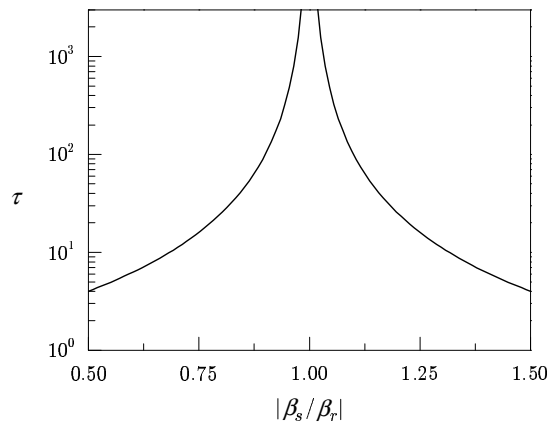}
\caption{\label{fig:fig2} The relaxation time $\tau=|{\rm Re}(\lambda_2)|^{-1}$ plotted as a function of
$|\beta_s/\beta_r|$. Note the singularity at resonance, $|\beta_s/\beta_r|=1$.}
\end{figure}

If the coupling between cavities is perfect ($\epsilon=1$) and the parameters of the subsystems are the same
($\beta_{r,1} = \beta_{r,2}=\beta_r$, $\beta_{s,1} = \beta_{s,2}=\beta_s$) then the steady state is the pure
state
\begin{equation}
\label{eq:ss}
\ket{\phi_{\rm ss}}=\frac{1}{\sqrt{|\beta_r|^2+|\beta_s|^2}}\left(\beta_r^* \ket{00}+
\beta_s^* \ket{11}\right),
\end{equation}
where we use the abbreviated notation $\ket{00}\equiv\ket{0}_1\ket{0}_2$ and $\ket{11}
\equiv\ket{1}_1\ket{1}_2$. Then when $\beta_r = \beta_s$, which we shall refer to as the \textit{resonance}
condition, the steady state is a maximally-entangled Bell state. This may seem to be ideal, but a problem
arises when we consider the eigenvalues of the operator $\mathcal{L}$. Specifically, the characteristic
time for the system to reach steady state, $\tau=\left|{\rm Re}(\lambda_2)\right|^{-1}$, where $\lambda_2$
denotes the eigenvalue of $\mathcal{L}$ with smallest (in magnitude) non-zero real part, approaches infinity
as the resonance condition is approached. This is shown by the plot in Fig.~\ref{fig:fig2}. Thus the master
equation itself, in particular its steady state, offers limited insight into the behavior of the system at
resonance. We wish to learn more about this special case; in particular, how does the entanglement develop
dynamically. Also, if additional information is factored into the description, by making measurements on the
environment, can we better characterize the long term behavior, or possibly find perfect entanglement after
a finite time? We demonstrate that quantum trajectory theory can provide answers to these questions.

\section{Quantum Trajectories}
\label{subsec3_1}

As with any open system, the first step in unraveling the master equation is to identify the points of coupling
to the environment. The first is obvious -- the output from Cavity 2. To measure this output, let us assume
the existence of an ideal photon detector in the path of the output from Cavity 2; we call it \emph{Detector 1}.

\begin{figure}[t]
\begin{center}
\includegraphics[width=7cm]{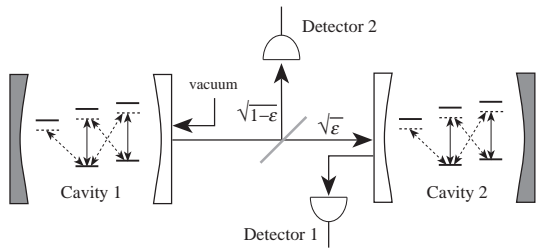}
\end{center}
\caption{Conceptual photon detectors, Detector 1 and Detector 2, used for unravelling the master equation. The
Faraday isolators are omitted for clarity.}
\label{fig:fig3} \end{figure}

The second point of coupling to the environment is more subtle. Our model does not assume the inter-cavity
coupling to be perfect; only a fraction $\epsilon$ of the output photon flux from Cavity 1 makes it into
Cavity 2. Physically, this loss may be caused, for example, by non-ideal transmissivity of the Faraday isolators
or by absorption in the cavity mirrors. These imperfections cause photons to be scattered into the environment
in some uncontrollable fashion.  Formally, though, this is equivalent to assuming that the apparatus is ideal,
except that there exists a beamsplitter between the cavities, as drawn schematically in Fig.~\ref{fig:fig3}.
We therefore further assume the existence of a second photon detector to collect photons reflected by this
beamsplitter; we call it \emph{Detector 2}. 

We now proceed to develop the quantum trajectory formalism for the cascaded qubit system. In this approach the
system is described by a pure state which is dependent on (conditioned on) the counting histories, or records,
of Detectors 1 and 2. Firstly, we rewrite the master equation in a form suitable for translation into the quantum
trajectory language. We reexpress Eq.~(\ref{eq:me}) in the form
\begin{eqnarray}
\dot{\rho}&=(\mathcal{L}_0 + \mathcal{S})\rho,
\end{eqnarray}
with
\begin{subequations}
\begin{align}
\mathcal{L}_0\rho&\equiv-i\left[\hat H_0,\rho\right]-\frac{1}{2}\sum_{i=1,2}
\left(\hat C_i^{\dag}\hat C_i \rho+\rho\hat C_i^{\dag}\hat C_i\right),\\
\mathcal{S}\rho&=\sum_{i=1,2}\hat C_i \rho\hat C_i^{\dag},
\end{align}
\end{subequations}
where
\begin{subequations}
\begin{eqnarray}
\hat C_1&=&\sqrt{2}\left(\sqrt{\epsilon}\hat R_1-\hat R_2\right),\\
\hat C_2&=&\sqrt{2(1-\epsilon)}\,\hat R_1,\\
\hat H_0&=&i\sqrt{\epsilon}\left(\hat R_2^{\dag}\hat R_1-\hat R_1^{\dag}\hat R_2\right).
\end{eqnarray}
\end{subequations}
Then, within quantum trajectory theory, the evolution of the system is described by a pure state $\ket{\phi}$
which evolves under the non-Hermitian effective Hamiltonian
\begin{equation}
\hat H_{\rm eff}=\hat H_0-i\frac{1}{2}\sum_{i=1,2}\hat C_i^{\dag}\hat C_i ,
\end{equation}
the continuous evolution interrupted at random times by quantum jumps, $\ket{\phi}\rightarrow\hat C_i\ket
{\phi}$, where the jumps occur with probability
\begin{eqnarray}
p_i(t) dt = \frac{\bra{\phi}\hat C_i^{\dag}\hat C_i\phik}{\langle\phi |\phi\rangle} dt
\end{eqnarray}
in time interval $(t,t+dt)$. Physically, the jump operators $\hat C_1$ and $\hat C_2$ account for the reduction
of the state of the system, given a photon count is recorded by Detector~1 or Detector~2, respectively. Thus,
within the quantum trajectory description of the coupled cavity system, we consider an experiment in which
ideal detectors are employed, such that every scattered photon is detected and recorded. Given the history of
detector `clicks', one has complete information about the system state, in the sense that that state is always
pure; hence, although the solution to the master equation is generally mixed, one is able to characterize the
entanglement in an unambiguous (conditional) fashion \cite{Nha04a}.

Consider the special case where the coupling between the cavities is optimal ($\epsilon=1$). In this case
there is only one output from the system, that from Cavity 2, recorded by Detector 1. Standard numerical
algorithms \cite{QOToolbox} have been used to simulate typical quantum trajectories for various values of
$|\beta_s/\beta_r|$. Specifically, we consider the evolution of the conditional expectation of the operator
product $\hat\sigma_{1,z}\hat\sigma_{2,z}$, where $\hat\sigma_{i,z}$ is the Pauli operator diagonal in the
$\left(\ket{0}_i,\ket{1}_i\right)$--representation,
\begin{equation}
\hat\sigma_{i,z}\ket{1}_i = \ket{1}_i, \qquad \hat\sigma_{i,z}\ket{0}_i =
-\ket{0}_i \, .
\end{equation}
This expectation has a number of convenient properties; for example, the steady-state value
\begin{equation}
\mean{\hat\sigma_{1,z}\hat\sigma_{2,z}}_{\rm ss}=1,
\end{equation}
regardless of the value of $|\beta_s/\beta_r|$, which makes it easy to compare rates of convergence to the
steady state for different system parameters.

\begin{figure}[b]
\begin{center}
\hbox{\hskip-0.08cm
\includegraphics[width=7.5cm]{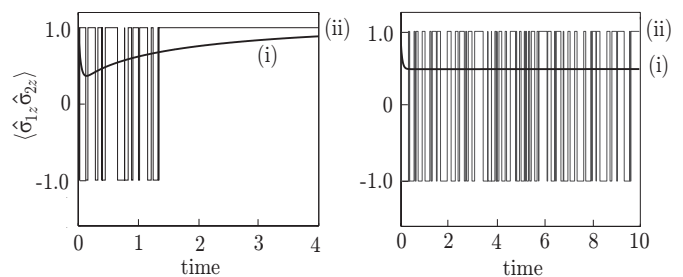}}
\vskip-0.5cm
\end{center}
\caption{Evolution of the ensemble average (i) compared with that of a single quantum trajectory (ii), for
$\epsilon=1$. The off-resonance case ($\beta_s\neq\beta_r$) is shown to the left and compared with the
resonant case ($\beta_s=\beta_r$) to the right. Time is measured in units of $\left(\beta_r/\sqrt\kappa
\right)^{-1}$.}
\label{fig:fig4}
\end{figure}

Figure~\ref{fig:fig4} contrasts the solution to the master equation and a single quantum trajectory. The
solution to the master equation exhibits a completely smooth evolution that tends asymptotically towards
the steady state. The quantum trajectory, on the other hand, undergoes a sequence of switches between two
extreme values of $\mean{\hat\sigma_{1,z}\hat\sigma_{2,z}}$, which occur at each photon detection. Provided
the parameters are chosen away from resonance, the photon detections eventually stop and the trajectory
settles into the steady state (\ref{eq:ss}), with $\mean{\hat\sigma_{1,z}\hat\sigma_{2,z}}=1$; the steady
state is clearly a dark state. At resonance, however, the photon detections may continue indefinitely.
Physically, this seems plausible, since it simply implies that the atoms continue to switch between states
$\ket{0}$ and $\ket{1}$, scattering one photon with each transition. At resonance, apparently, a unique
equilibrium dark state cannot be established. The cyclic behavior that replaces it is completely invisible
if we consider only the ensemble average--a vivid demonstration of how single quantum trajectories can provide
additional insight into the evolution of an open quantum system.

\section{Entangled-state Cycles}
\label{subsec4_1}

The oscillatory behavior featured in Fig.~\ref{fig:fig4} hints at a simple cyclic process. In fact, it
is simple enough that we can understand why it occurs without resorting to numerics. In this section we
formulate a graphical description of individual trajectories.

\subsection{The Cascaded System Phase Space}

Figure~\ref{fig:fig4} demonstrates that the conditional expectation $\mean{\hat\sigma_{1,z}\hat\sigma_{2,z}}$
is conserved during the periods of evolution between quantum jumps. The positively and negatively correlated
subspaces
\begin{equation}
E^{\pm}=\left\{\ket{\phi}:\mean{\hat\sigma_{1,z}\hat\sigma_{2,z}}=\pm 1\right\}
\end{equation}
are coupled only through quantum jumps. Noting that
\begin{equation}
E^+ = \textrm{span}\{\ket{00}, \ket{11}\}, \quad E^-=\textrm{span}\{\ket{10},\ket{01}\}
\end{equation}
are each $2$-dimensional (assuming real amplitudes without loss of generality), we manage to break up
a $4$-dimensional space into two $2$-dimensional planes, linked to one another by the quantum jumps.
We refer to this representation as the \textit{cascaded system phase space}. Trajectories within it can
be viewed as lines moving continuosly within either plane and jumping discontinuously between the planes.

\subsection{Evolution Between Quantum Jumps}

We use phase space portraits within $E^+$ and $E^-$ to characterize the behavior of the system, where for
the sake of simplicity, and without loss of generality, we are assuming $\beta_r$ and $\beta_s$ to be real.
We define
\begin{equation}
r =|\beta_s/\beta_r|=\beta_s/\beta_r,
\end{equation}
and scale time by setting $\beta_r/\sqrt{\kappa} = 1$. The master equation then takes the form ($\epsilon=1$)
\begin{align}\label{eqn:meq}\nonumber
\dot{\rho}=&\sum_{i=1,2}\left(2\hat R_i\rho\hat R_i^{\dag}-\hat R_i^{\dag}\hat R_i\rho-
\rho\hat R_i^{\dag}\hat R_i\right)\\
&+2(\rho\hat R_1^{\dag}\hat R_2-\hat R_2\rho\hat R_1^{\dag}+\hat R_2^{\dag}\hat R_1 \rho-
\hat R_1\rho\hat R_2^{\dag}),
\end{align}
where
\begin{equation}
\hat R_i=\hat\sigma_i^-+r\hat\sigma_i^+.
\end{equation}
The resonance condition is now $r=1$.

It is useful to convert to a matrix notation, such that a pure state $\ket{\phi}$ of the system is represented
by a $4$-vector,
\begin{equation}
\ket{\phi} = (c_{11},c_{10},c_{01},c_{00})^T \equiv \sum_{i,j = 0,1}
c_{ij}\ket{ij} \, ,
\end{equation}
and system operators are written as $4\times4$ matrices, e.g.,
\begin{eqnarray}
\hat R_1=\left(\begin{array}{cccc}
0&0&r&0\\
0&0&0&r\\
1&0&0&0\\
0&1&0&0\\\end{array}\right),\quad
\hat R_2=\left(\begin{array}{cccc}
0&r&0&0\\
1&0&0&0\\
0&0&0&r\\
0&0&1&0\\\end{array} \right) ,
\end{eqnarray}
and
\begin{eqnarray}
\hat C_1&=&\sqrt{2}\left(\begin{array}{cccc}
0&-r&r&0\\
-1&0&0&r\\
1&0&0&-r\\
0&1&-1&0\\\end{array}\right) .
\end{eqnarray}
The evolution of $\ket{\phi}$under $\hat H_{\rm eff}$ is written as a linear differential equation in four
variables,
\begin{align}
\ddt\ket{\phi}&=-i\hat H_{\rm eff}\phik=\left[-i\hat H_0-\half\hat C_1^{\dag}\hat C_1\right]\! 
\ket{\phi}\nonumber\\
&=\left(\begin{array}{cccc}
-2&0&0&2r\\
0&-(1+r^2)&2r^2&0\\
0&2&-(1+r^2)&0\\
2r&0&0&-2r^2\\\end{array}\right)
\label{eqn:basicde}\!\ket{\phi}.
\end{align}
As noted above, this evolution is constrained within either $E^+$ or $E^-$. Thus we can write $\ket{\phi}$ as
a vector sum of two orthogonal components $\ket{\phi^+}\in E^+$ and $\ket{\phi^-}\in E^-$, $\ket{\phi}=a\ket
{\phi^+}+b\ket{\phi^-}$, to obtain the decoupled dynamics
\begin{subequations}
\begin{eqnarray}
\ddt\ket{\phi^+}&=&\left(\begin{array}{cc}
-2&2r\\
2r&-2r^2\\
\end{array}\right)
\ket{\phi^+},\\
\ddt\ket{\phi^-}&=&\left(\begin{array}{cc}
-(1+r^2)&2r^2\\
2&-(1+r^2)\\
\end{array}\right)
\ket{\phi^-}.
\end{eqnarray}
\end{subequations}
Eigenvectors of the two dynamical matrices correspond to states of the system that are preserved under the
evolution between quantum jumps. Note, however, that it does not necessarily follow that such a state is a
steady state of the quantum trajectory evolution as a whole; it must eventually experience a quantum jump
if its norm decays---i.e., the corresponding eigenvalue is not zero. Recall from quantum trajectory theory
that the probability for a state not to jump prior to time $t$ is given by its norm \cite{Carmichael93}.

For the systems of equations given above we find the following (unnormalised) eigenstates and
eigenvalues:
\begin{enumerate}
\item[(i)]
$\ket{\phi_1} = \ket{00} + r \ket{11}$, ~$\lambda_1 = 0$; this is the steady state of the system
for $r<1$. 
\item[(ii)]
$\ket{\phi_2} = - r\ket{00} + \ket{11}$, ~$\lambda_2 = -2(1+r^2)$; this state in $E^+$ is orthogonal
to $\ket{\phi_1}$ and must eventually jump to a state in $E^-$.
\item[(iii)]
$\ket{\phi_3} = r\ket{10} + \ket{01}$, ~$\lambda_3 = -(r-1)^2$; this state in $E^-$ must eventually
jump to a state in $E^+$ unless $r=1$; in the latter case it plays no role once an entangled-state
cycle is initiated (see below).
\item[(iv)]
$\ket{\phi_4} = \ket{10} - r \ket{01}$, ~$\lambda_4 = -(r+1)^2$; this state in $E^-$ must eventually
jump to a state in $E^+$.
\end{enumerate}

In the special case of resonance, $r=1$, there are two independent steady states, $\ket{\phi_1}$
and $\ket{\phi_3}$, which helps to explain the failure of the master equation evolution to
approach a unique steady state. It also suggests a fundamental feature of the indefinite switching,
the cyclic behavior, revealed by individual quantum trajectories: during such an \emph{entangled-state
cycle}, the system state must remain orthogonal to $\ket{\phi_1}$ and $\ket{\phi_3}$.
We verify this shortly, after examining the trajectory evolution away from resonance, where the steady
state $\ket{\phi_1}$ is always reached for perfect inter-cavity coupling.

\subsection{Quantum Trajectories for $r<1$}

Typical quantum trajectories for $r=0.5$ are shown in Figs.~\ref{fig:fig5} and
\ref{fig:fig6}, where the $E^+$ and $E^-$ subspaces are drawn as circular planes. Normalized
states are located on the circumferences of the circles. The Bell states
\begin{subequations}
\begin{eqnarray}
\ket{\Phi^\pm}&=&(\ket{00}\pm\ket{11})/\sqrt{2},\\
\ket{\Psi^\pm}&=&(\ket{01}\pm\ket{10})/\sqrt{2},
\end{eqnarray}
\end{subequations}
lie at intersections of the circumference with the dotted lines as shown.

Between quantum jumps, under the influence of the non-Hermitian Hamiltonian $\hat H_{\rm eff}$, the
norm of the state decays and the point representing it within the phase space moves to the interior
of one of the circles. Quantum jumps cause a switch from $E^+$ to $E^-$ or vice-versa. They are
represented by the lines connecting the two planes, where for illustrative purposes, the system
state is renormalized after each quantum jump; thus jumps terminate at points on the circumference
of the circles. 

We restrict ourselves to separable initial states located in one or other of the two subspaces;
for example, the states $\ket{00}$ and $\ket{10}$, respectively, are considered in Figs.~\ref{fig:fig5}
and \ref{fig:fig6}. 

\subsubsection{Effect of quantum jumps}

The action of the jump operator $\hat C_1$ on states located in $E^+$ (with renormalization) is
\begin{subequations} 
\begin{equation}
\hat C_1\left(\begin{array}{c}
c_{11}\\0\\0\\c_{00}\end{array}\right)\rightarrow 
\frac{\textrm{sign}\{c_{11}-rc_{00}\}}{\sqrt{2}}
\left(\begin{array}{c}
0\\-1\\1\\0\end{array}\right),
\end{equation}
while the action of $\hat C_1$ on states in $E^-$ is
\begin{equation}
\hat C_1 \left(\begin{array}{c}
0\\c_{10}\\c_{01}\\0\end{array}\right)\rightarrow
\frac{\textrm{sign}\{ c_{10}-c_{01}\}}{\sqrt{1+r^2}}
\left(\begin{array}{c}
-r\\0\\0\\1\end{array}\right) .
\end{equation}
Thus, when a quantum jump occurs, any state within $E^+$ collapses onto the Bell state $\ket{\phi}
=\pm \ket{\Psi^-}$ in $E^-$, while any state within $E^-$ collapses onto the state $\ket{\phi}
=\pm (\ket{00}-r\ket{11})/\sqrt{1+r^2}$ in $E^+$.
\end{subequations}

\subsubsection{Jump probabilities}

Consider an initial normalized state in $E^+$,  $\ket{\phi^+(0)}=a\ket{\phi_1}+b\ket{\phi_2}$, for some
(real) $\{ a,b\}$. Given that $\ket{\phi_1}$ is a steady state of the evolution between quantum jumps,
the probability of an eventual quantum jump to $E^-$ is
\begin{eqnarray}
P_{+\rightarrow-} = \left|\langle\phi^+(0)|\phi_2\rangle\right|^2 = b^2 ,
\end{eqnarray}
while with probability $\left|\langle\phi^+(0)|\phi_1\rangle\right|^2=1-P_{+\to-}=a^2$ the system evolves
to the steady state $\ket{\phi_1}$ without any photon emissions.

If a jump from $E^-$ to $\ket{\phi}=\pm (\ket{00}-r\ket{11})/\sqrt{1+r^2}$ has just occurred, then by the
same argument one shows that the probability of a future quantum jump to $E^-$ is $4r^2/(1+r^2)^2$, or,
alternatively, the probability of reaching the steady state after such a jump is $1-4r^2/(1+r^2)^2=[(1-r^2)
/(1+r^2)]^2$.

\begin{figure}[t]
\begin{center}
\vbox{
\includegraphics[scale=0.4]{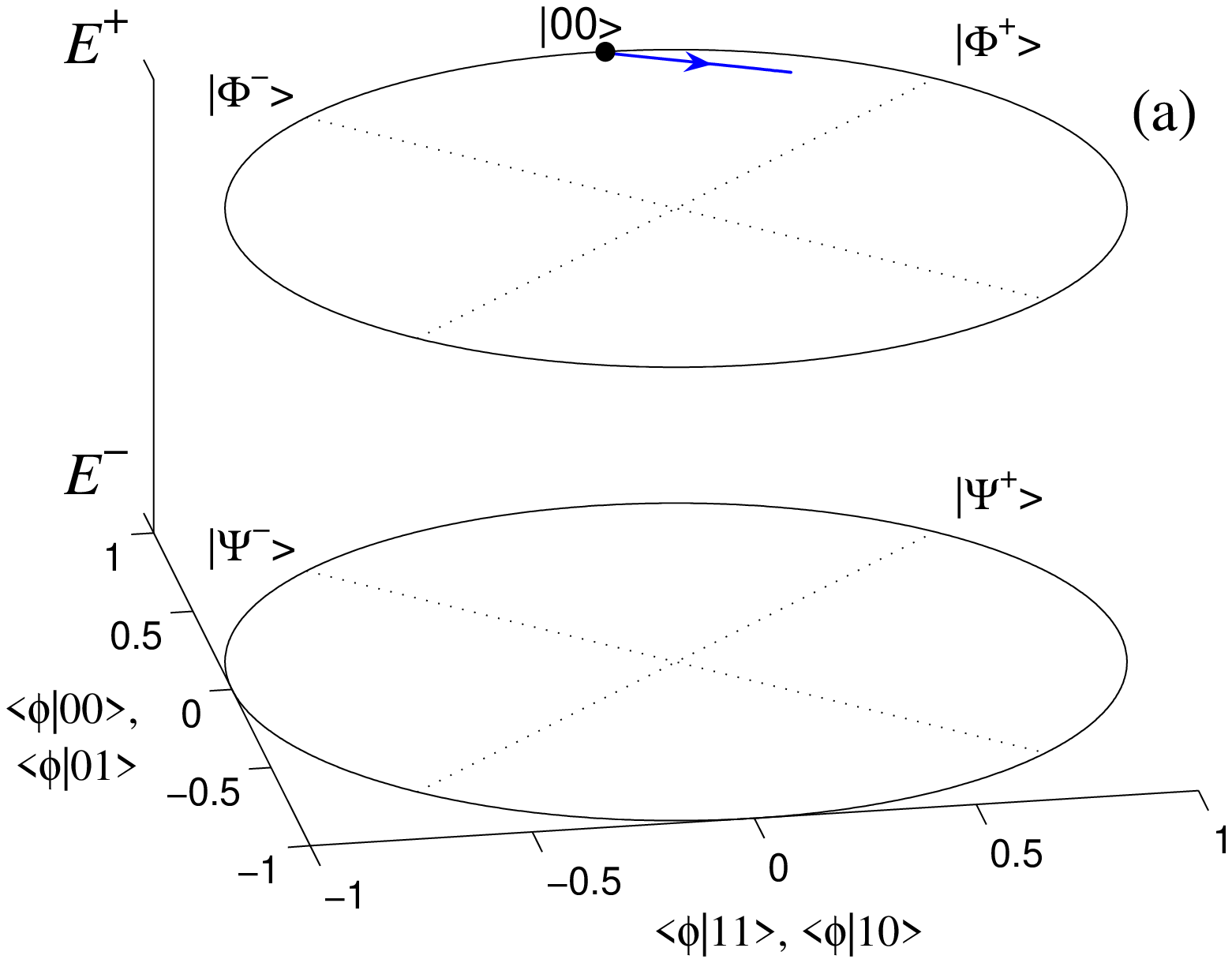}
\vskip0.5cm
\includegraphics[scale=0.4]{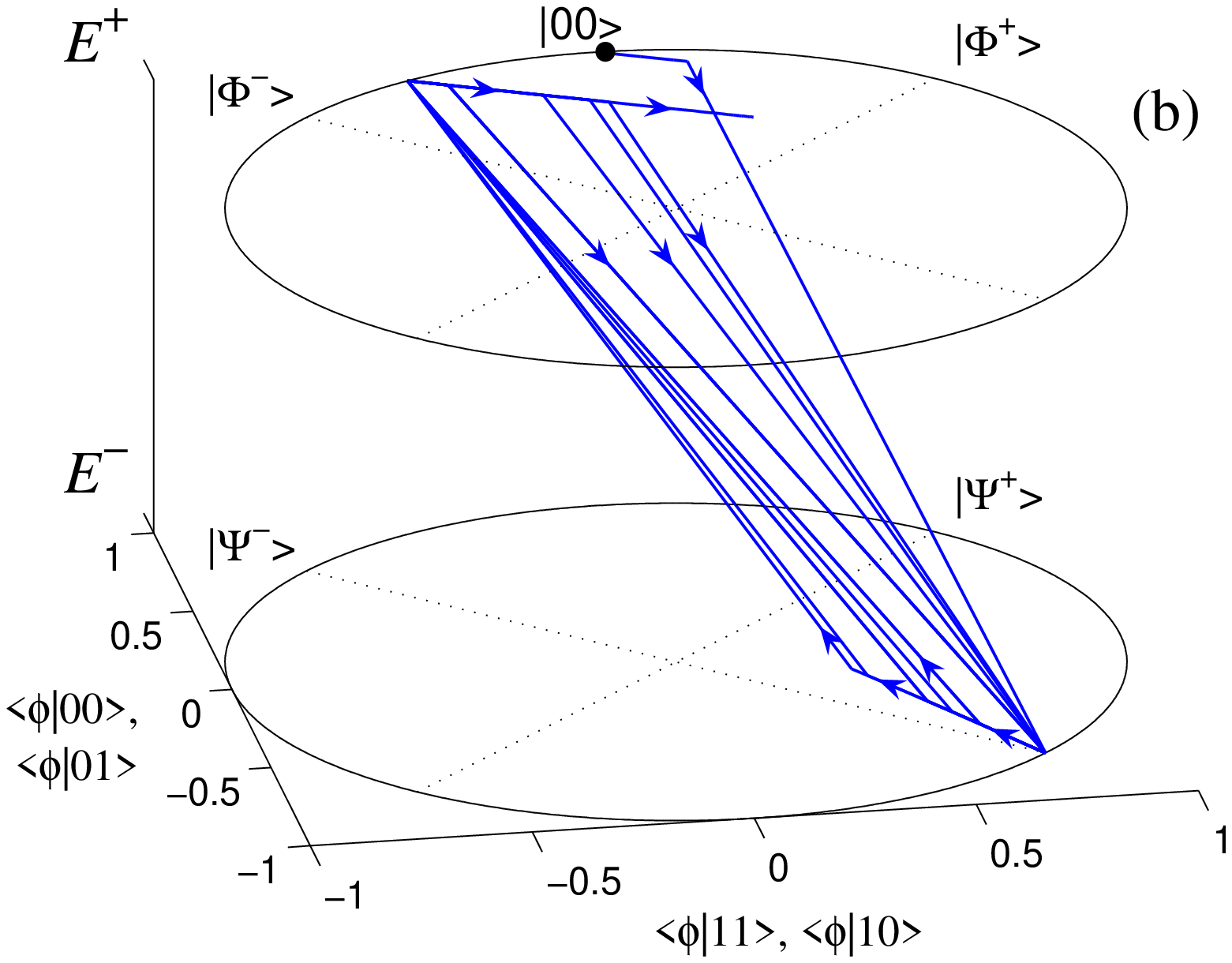}
\vskip0.5cm
\includegraphics[scale=0.4]{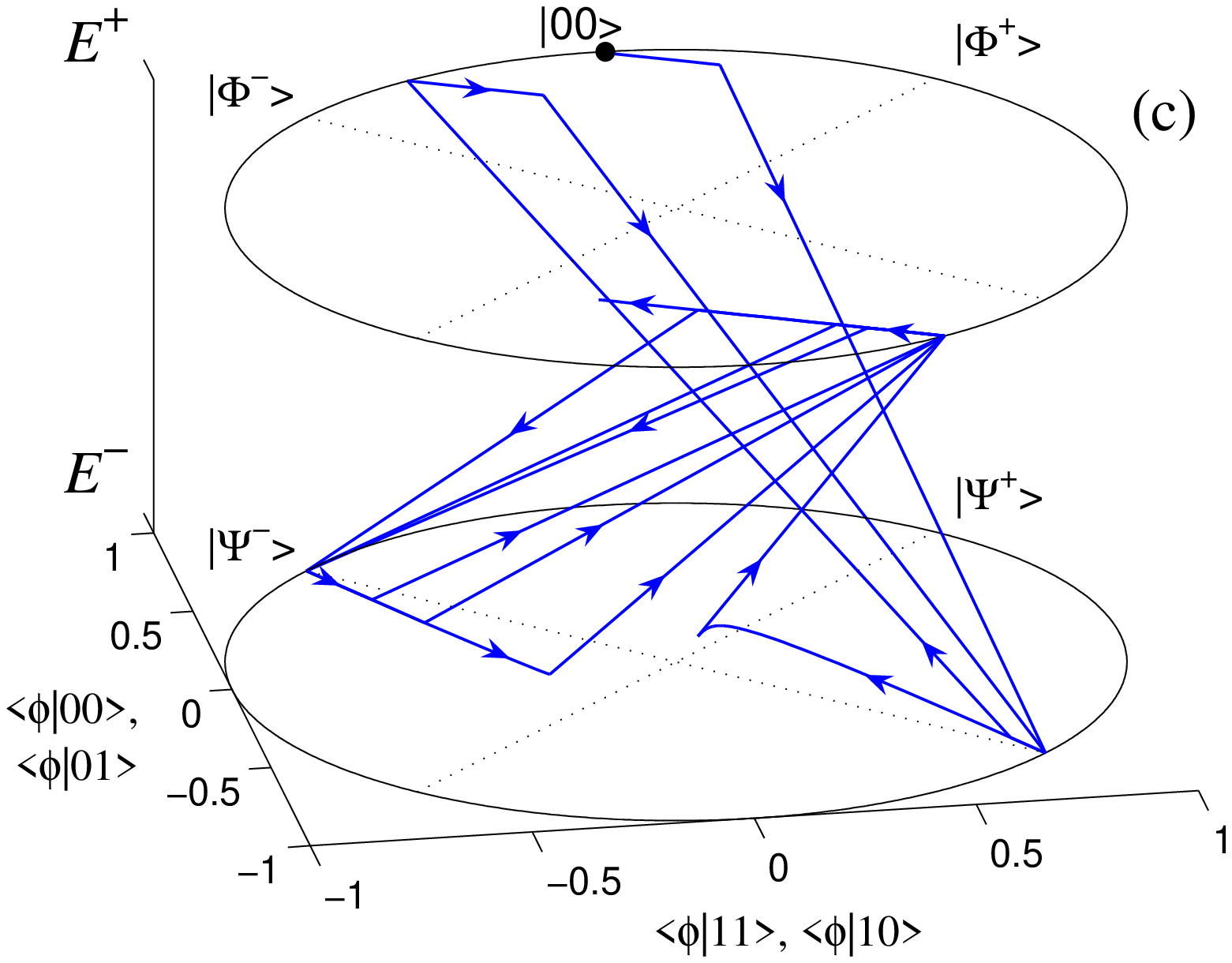}
}
\caption{
Examples of phase space trajectories for $r=0.5$ with initial state $\ket{\phi^+(0)}=\ket{00}$ and
$\epsilon=1$. See text for description.
}
\label{fig:fig5}
\end{center}
\end{figure}

\begin{figure}[t]
\begin{center}
\vbox{
\includegraphics[scale=0.4]{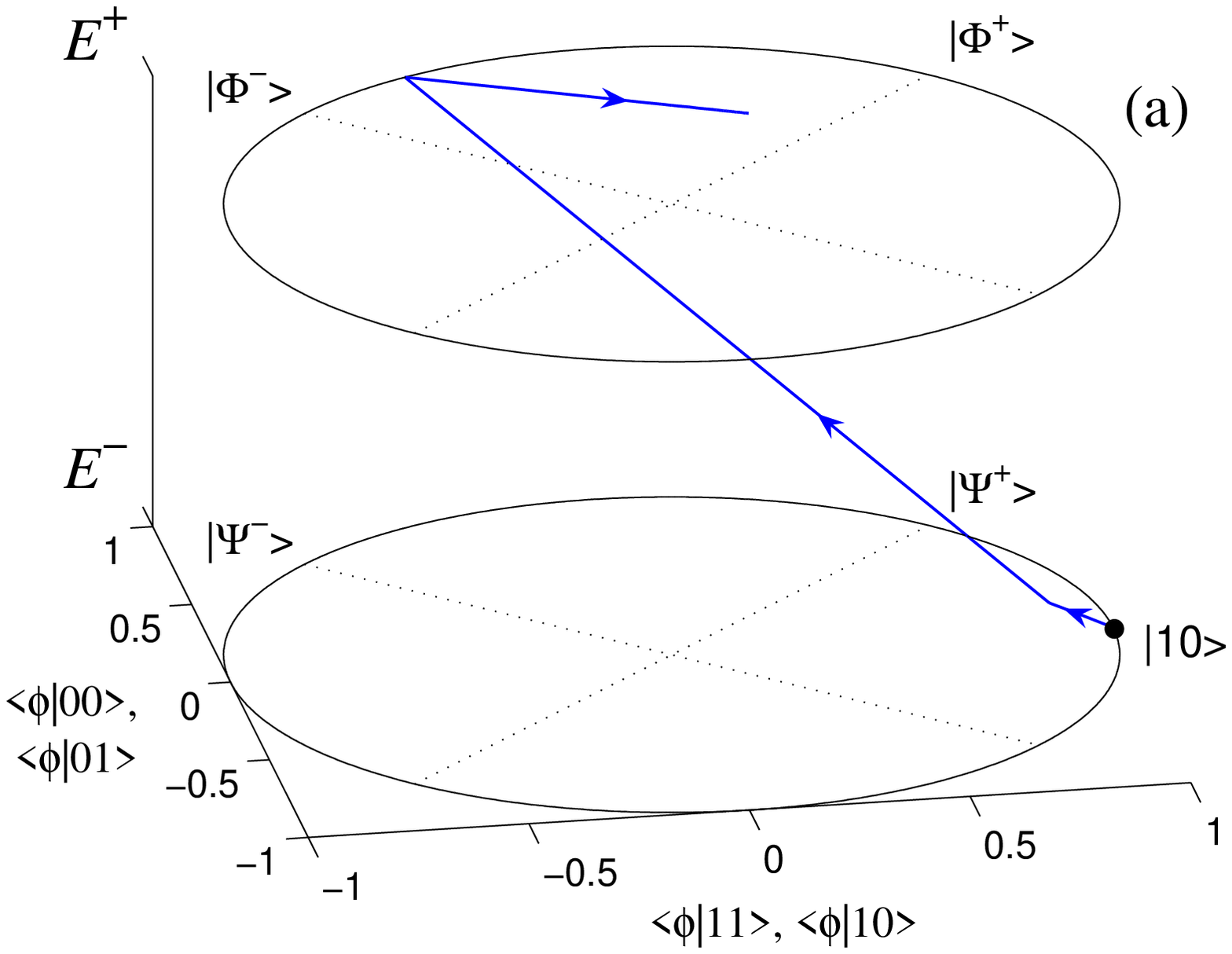}
\vskip0.5cm
\includegraphics[scale=0.4]{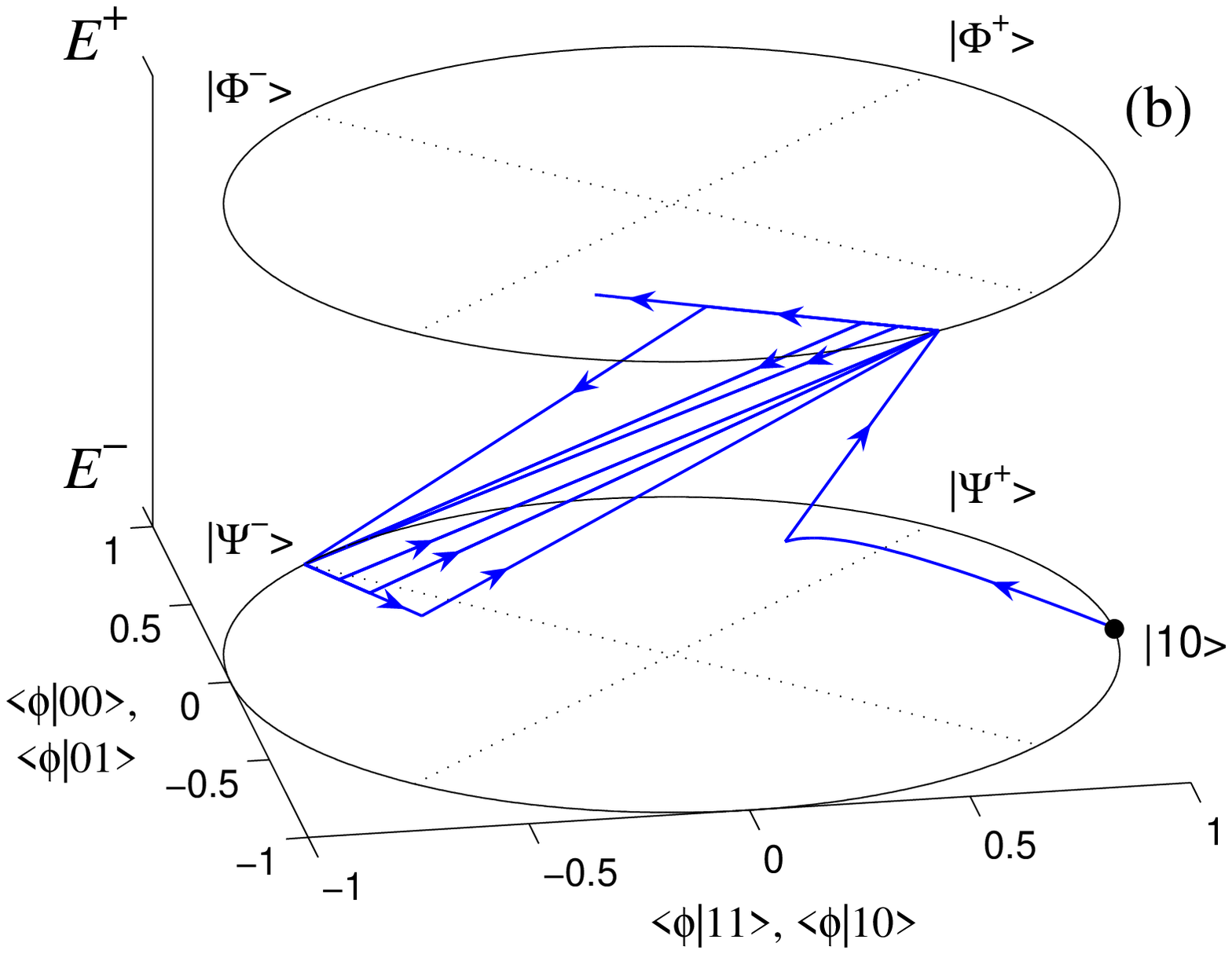}
\vskip0.5cm
\includegraphics[scale=0.4]{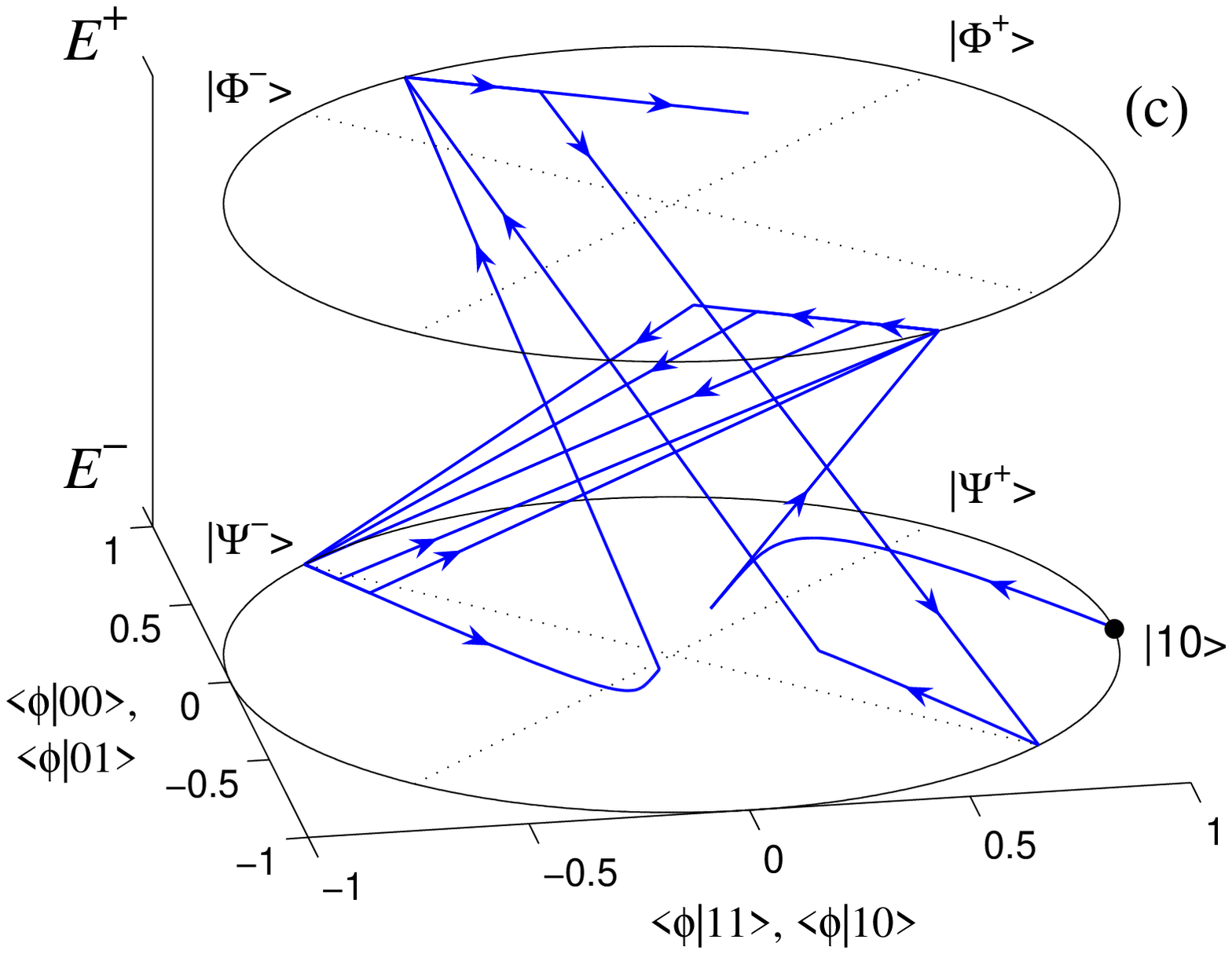}
}
\caption{
Examples of phase space trajectories for $r=0.5$ with initial state $\ket{\phi^-(0)}=\ket{10}$ and
$\epsilon=1$. See text for description.
}
\label{fig:fig6}
\end{center}
\end{figure}

Consider now an initial state in $E^-$, $\ket{\phi^-(0)}=c\ket{\phi_3}+d\ket{\phi_4}$, for some (real)
$\{ c,d\}$. Owing to the instability of both $\ket{\phi_3}$ and $\ket{\phi_4}$ for $r<1$, an eventual
quantum jump is guaranteed; thus,
\begin{eqnarray}
P_{-\rightarrow+} = 1.
\end{eqnarray}
Armed with this information, we move to an explanation of the quantum trajectories displayed in
Figs.~\ref{fig:fig5} and \ref{fig:fig6}. 

\subsubsection{Initial states $\ket{00}$ and $\ket{10}$}

In Fig.~\ref{fig:fig5} we plot three typical phase-space trajectories for $r=0.5$ and $\ket{\phi^+(0)}
=\ket{00}$. Fig.~\ref{fig:fig5}(a) illustrates the case where the system evolves directly to the steady
state $\ket{\phi_1}$. The probability of this event is $\left|\langle 00|\phi_1\rangle\right|^2=1/(1+r^2)
=0.8$, so it is the most likely occurrence for the chosen parameters. If a first quantum jump does occur,
then typical trajectories are shown in Figs.~\ref{fig:fig5}(b) and (c). Following the jump to $\ket{\phi}
=-\ket{\Psi^-}$ in $E^-$, a second jump returning the state to $E^+$ is guaranteed. For $r=0.5$, this
leaves the system in the state $\ket{\phi}=0.89\ket{00}-0.45\ket{11}$, from which the probability of a
further cycle of jumps is $4r^2/(1+r^2)^2=0.64$. Thus, after a first quantum jump cycle, it is most likely
that further cycles will follow, as seen in Figs.~\ref{fig:fig5}(b) and (c), where in both cases a total
of five cycles (ten photon detections) occur before the system finally reaches the steady state.

In Fig.~\ref{fig:fig6} we plot three typical phase-space trajectories for $r=0.5$ and $\ket{\phi^-(0)}
=\ket{10}$. In this case, at least one quantum jump is certain to occur, following which the probability
of further jumps is $4r^2/(1+r^2)^2=0.64$, as above. So for this initial condition, the most likely outcome
is a sequence of quantum jump cycles following a first guaranteed photon detection. In Fig.~\ref{fig:fig6}
(a) only the first detection occurs, while in Figs.~\ref{fig:fig6}(b) and (c) this detection is followed by
a sequence of cycles before the steady state is eventually achieved.

\subsection{Quantum Trajectories for $r=1$}

The case $r=1$ is of particular interest. The normalized eigenstates of the evolution between quantum jumps
are the Bell states $\ket{\phi_1}=\ket{\Phi^+}$, $\ket{\phi_2}=\ket{\Phi^-}$, $\ket{\phi_3}=\ket{\Psi^+}$,
and $\ket{\phi_4}=\ket{\Psi^-}$. The eigenvalues are $\lambda_1=\lambda_3=0$ and $\lambda_2=\lambda_4=-4$.
The action of the jump operator $\hat C_1$ on states within $E^+$ simplifies to 
\begin{subequations}
\begin{equation}
\mkern-45mu\hat C_1\left(\begin{array}{c}
c_{11}\\0\\0\\c_{00}\end{array}\right)\rightarrow 
\textrm{sign}\{c_{11}-c_{00}\}\ket{\Psi^-},
\label{eq:plusjump}
\end{equation}
and its action on states within $E^-$ to
\begin{equation}
\hat C_1\left(\begin{array}{c}
0\\c_{10}\\c_{01}\\0\end{array}\right)\rightarrow
\textrm{sign}\{c_{10}-c_{01}\}\ket{\Phi^-}.
\label{eq:minusjump}
\end{equation}
For $r=1$, photon detections, if they occur, are associated with collapses onto one of two maximally-entangled
Bell states.
\end{subequations}

For initial states $\ket{\phi^+(0)}$ and $\ket{\phi^-(0)}$ in $E^+$ and $E^-$, respectively, the system
evolves continuously, without the emission of any photons, to $\ket{\phi_1}=\ket{\Phi^+}$ and $\ket{\phi_3}
=\ket{\Psi^+}$, with probabilities $\left|\langle\phi^+(0)|\Phi^+\rangle\right|^2$ and $\left|\langle\phi^-(0)
|\Psi^+\rangle\right|^2$. Alternatively, a photon is detected with associated quantum jump to $\ket{\Psi^-}$
in $E^-$ or $\ket{\Phi^-}$ in $E^+$. In this case, as both terminal states are unstable under the between-jump
evolution, a second detection and quantum jump must follow. According to Eqs.~(\ref{eq:plusjump}) and
(\ref{eq:minusjump}) this simply exchanges $\ket{\Psi^-}$ for $\ket{\Phi^-}$ and vice-versa. Hence, a perpetual
switching between Bell states $\ket{\Psi^-}$ and $\ket{\Phi^-}$ occurs. We designate this behavior an
{\em entangled-state cycle}.

Thus, at resonance we find a distinctly bimodal behavior. The system either evolves into a maximally-entangled
Bell state without emitting photons, or an entangled-state cycle is initiated under which the system switches
indefinitely between orthogonal Bell states while emitting a continual stream of photons. As an aside, such
behavior can be regarded as a quantum measurement that distinguishes the Bell states $\ket{\Phi^\pm}$ from
$\ket{\Psi^\pm}$. 

\begin{figure}[b]
\begin{center}
\vbox{
\includegraphics[scale=0.4]{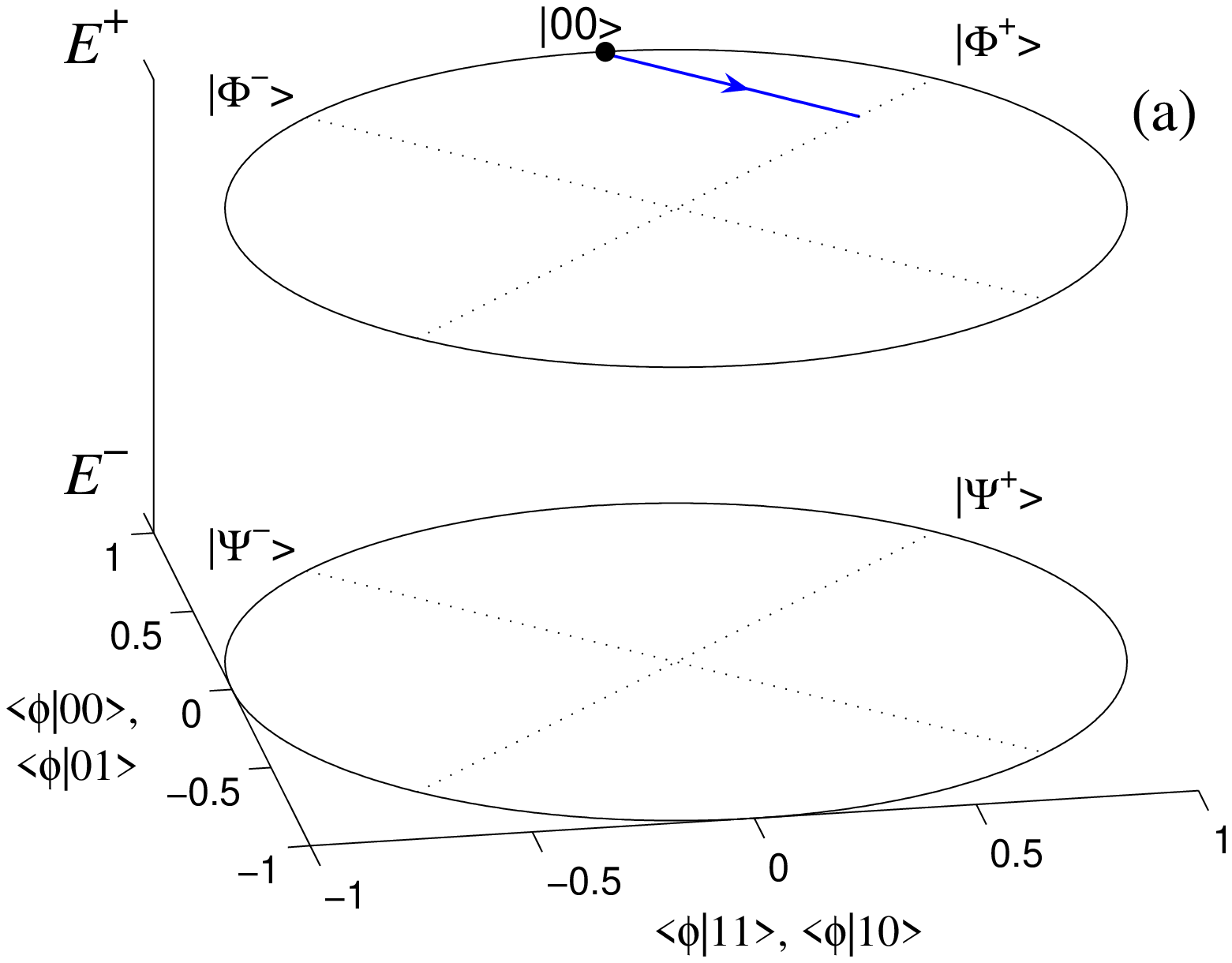}
\vskip0.5cm
\includegraphics[scale=0.4]{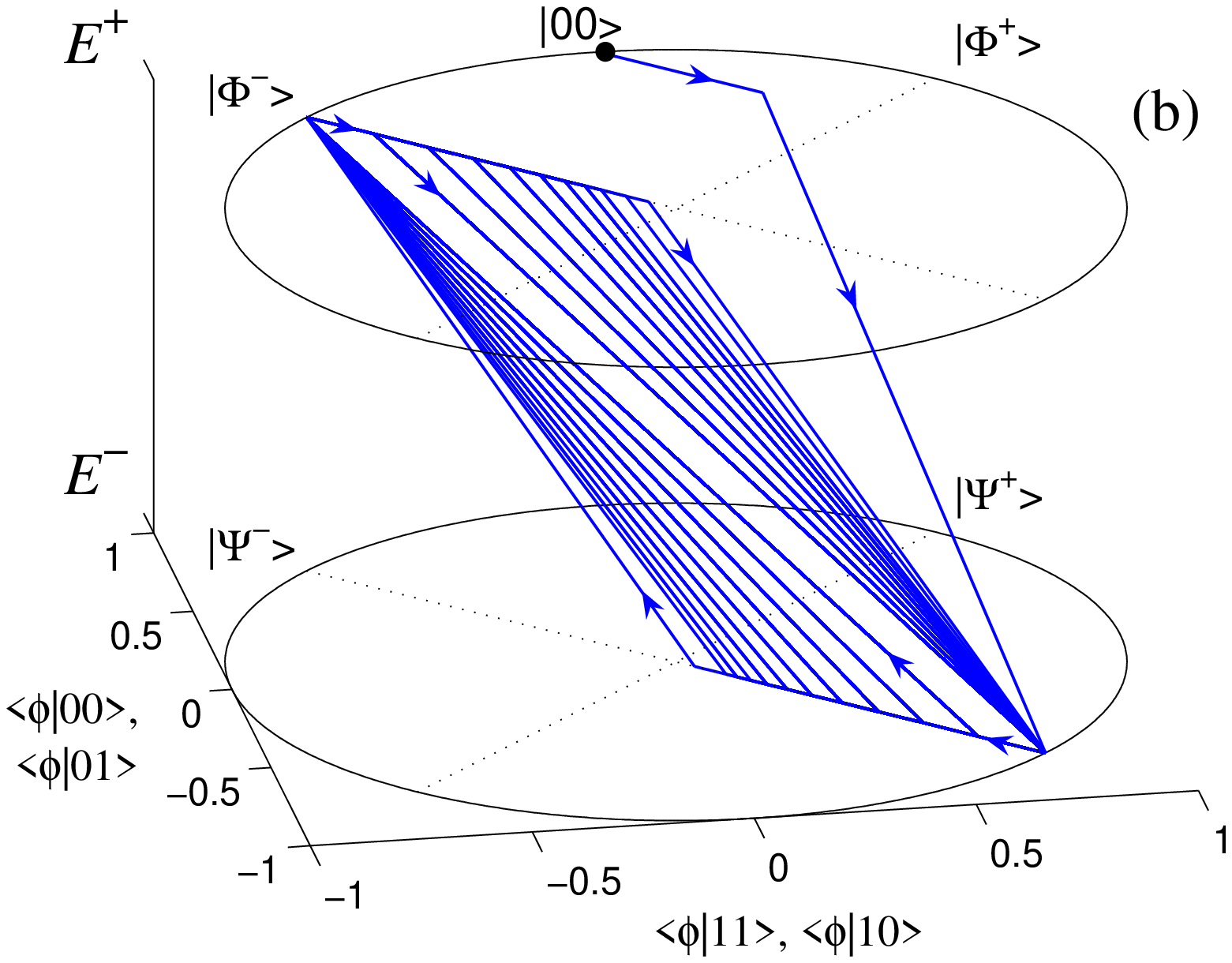}
}
\caption{
Examples of phase space trajectories for $r=1$ with initial state $\ket{\phi^+(0)}=\ket{00}$ and $\epsilon=1$.
See text for description.
}\label{fig:fig7}
\end{center}
\end{figure}

\begin{figure}[b]
\begin{center}
\vbox{
\includegraphics[scale=0.4]{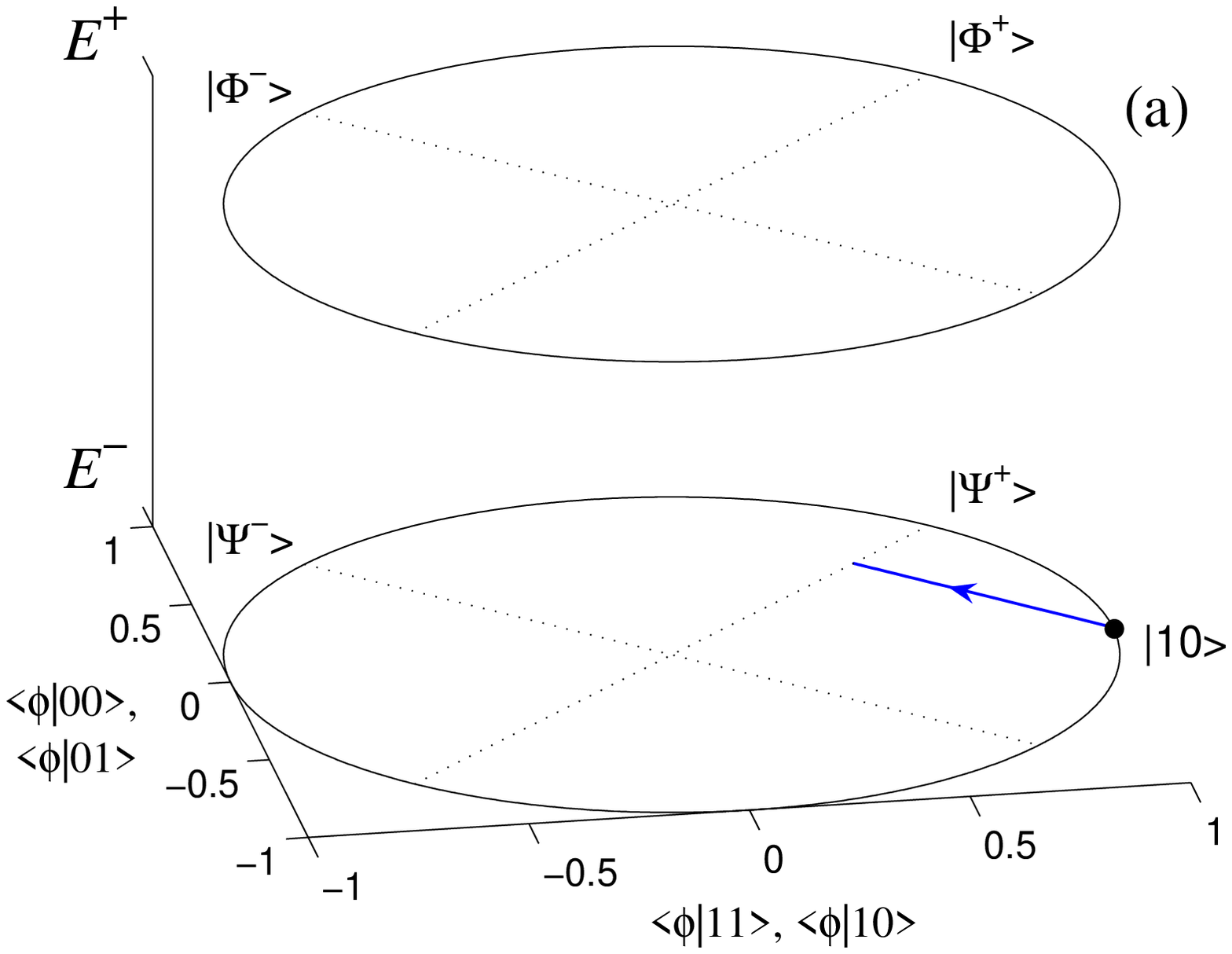}
\vskip0.5cm
\includegraphics[scale=0.4]{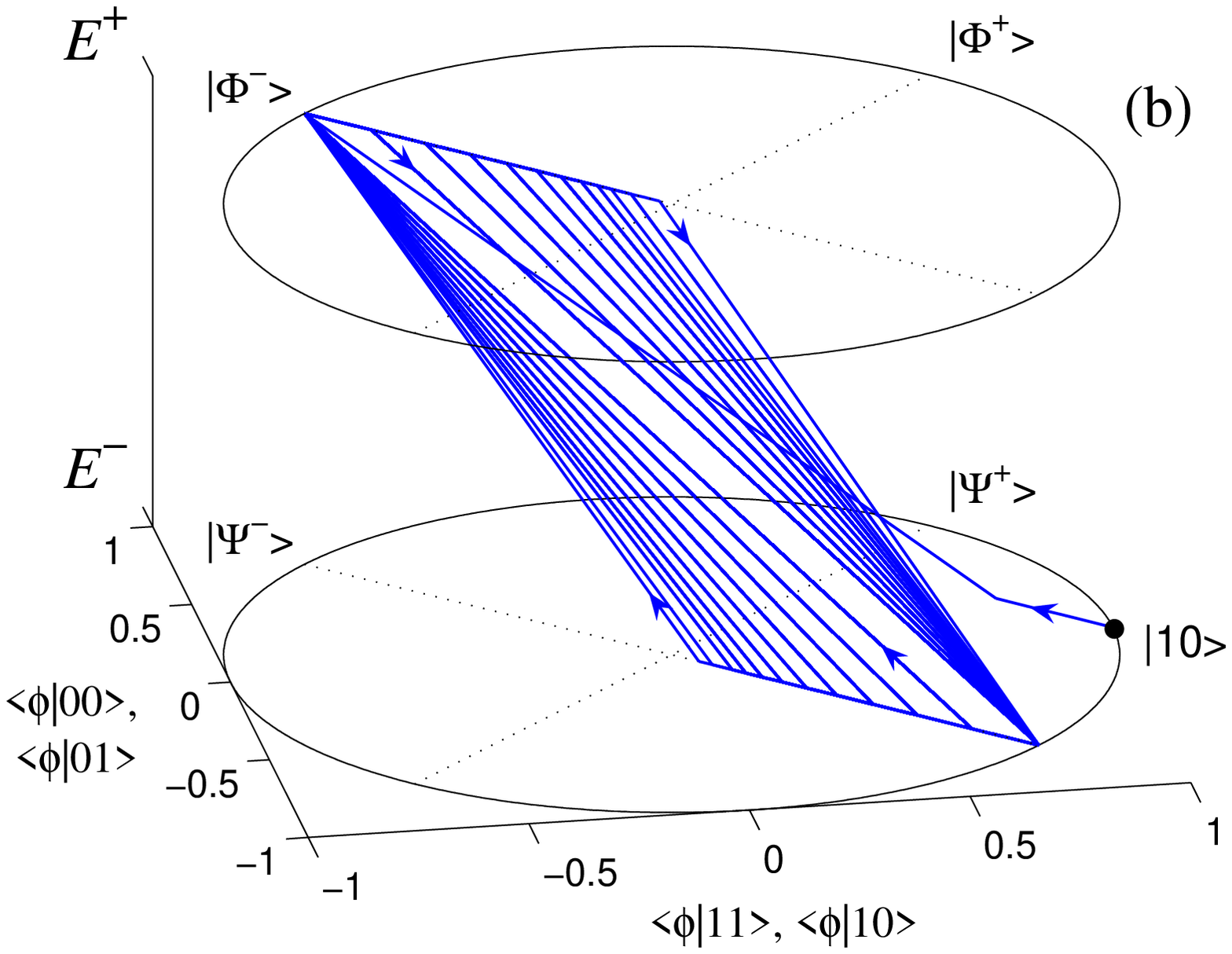}
}
\caption{
Examples of phase space trajectories for $r=1$ with initial state $\ket{\phi^-(0)}=\ket{10}$ and $\epsilon=1$.
See text for description.
}\label{fig:fig8}
\end{center}
\end{figure}

The two alternative outcomes of the quantum trajectory evolution are illustrated in Figs.~\ref{fig:fig7}
and \ref{fig:fig8} for the initial states $\ket{\phi^+(0)}=\ket{00}$ in $E^+$ and $\ket{\phi^-(0)}
=\ket{10}$ in $E^-$, respectively. With this choice of initial states there are equal probabilities for
reaching the steady states, $\ket{\Phi^+}$ [Fig.~\ref{fig:fig7}(a)] and $\ket{\Psi^+}$ [Fig.~\ref{fig:fig8}(a)],
and for commencing an entangled-state cycle [Figs.~\ref{fig:fig7}(b) and \ref{fig:fig8}(b)]. Note that once
an entangled-state cycle is initiated, the trajectory remains in a plane orthogonal to the lines defining
$\ket{\Phi^+}$ and $\ket{\Psi^+}$; the cycle continues indefinitely

\subsection{Imperfect Intercavity Coupling}

\begin{figure}[t]
\begin{center}
\vbox{
\includegraphics[scale=0.4]{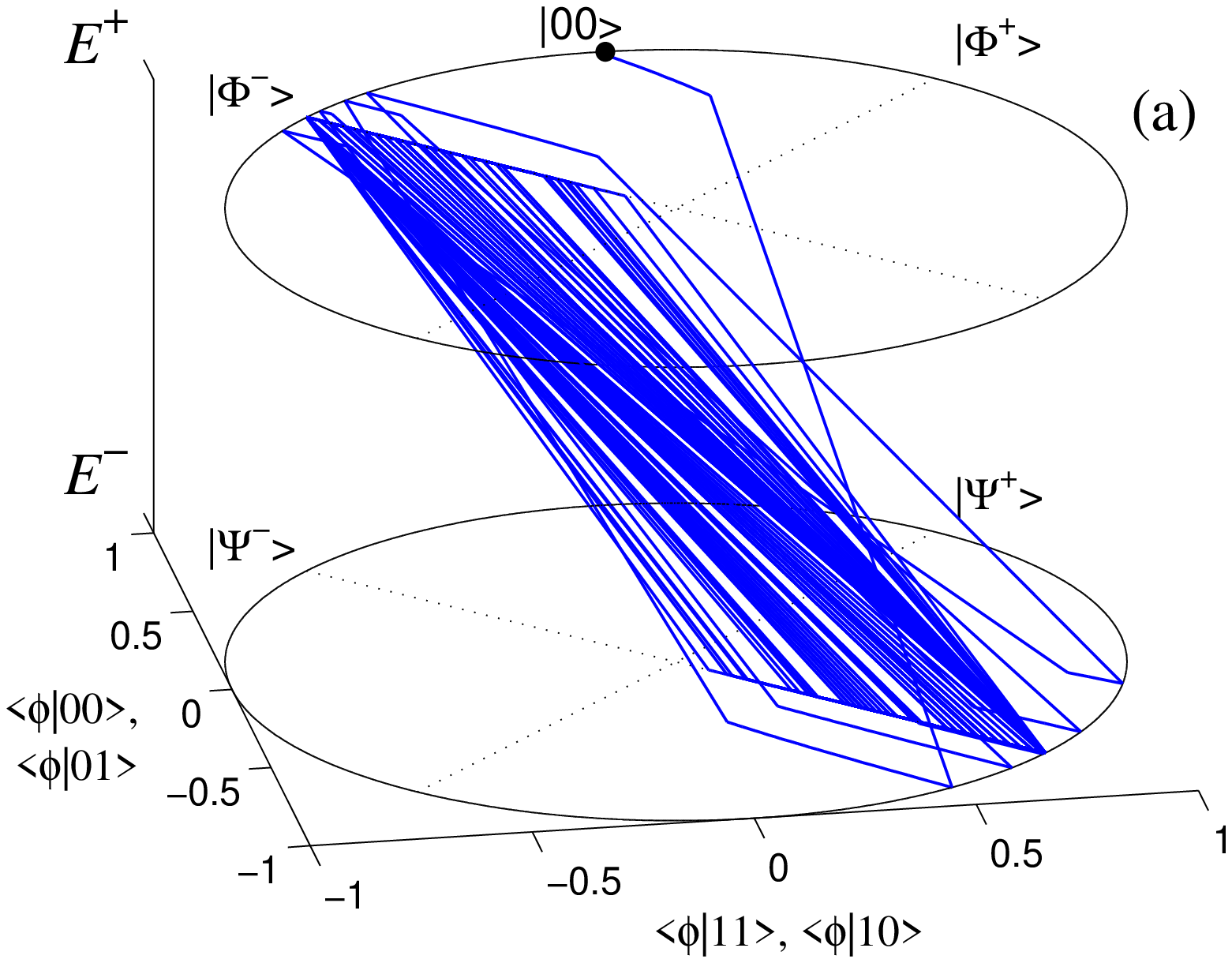}
\vskip0.5cm
\includegraphics[scale=0.4]{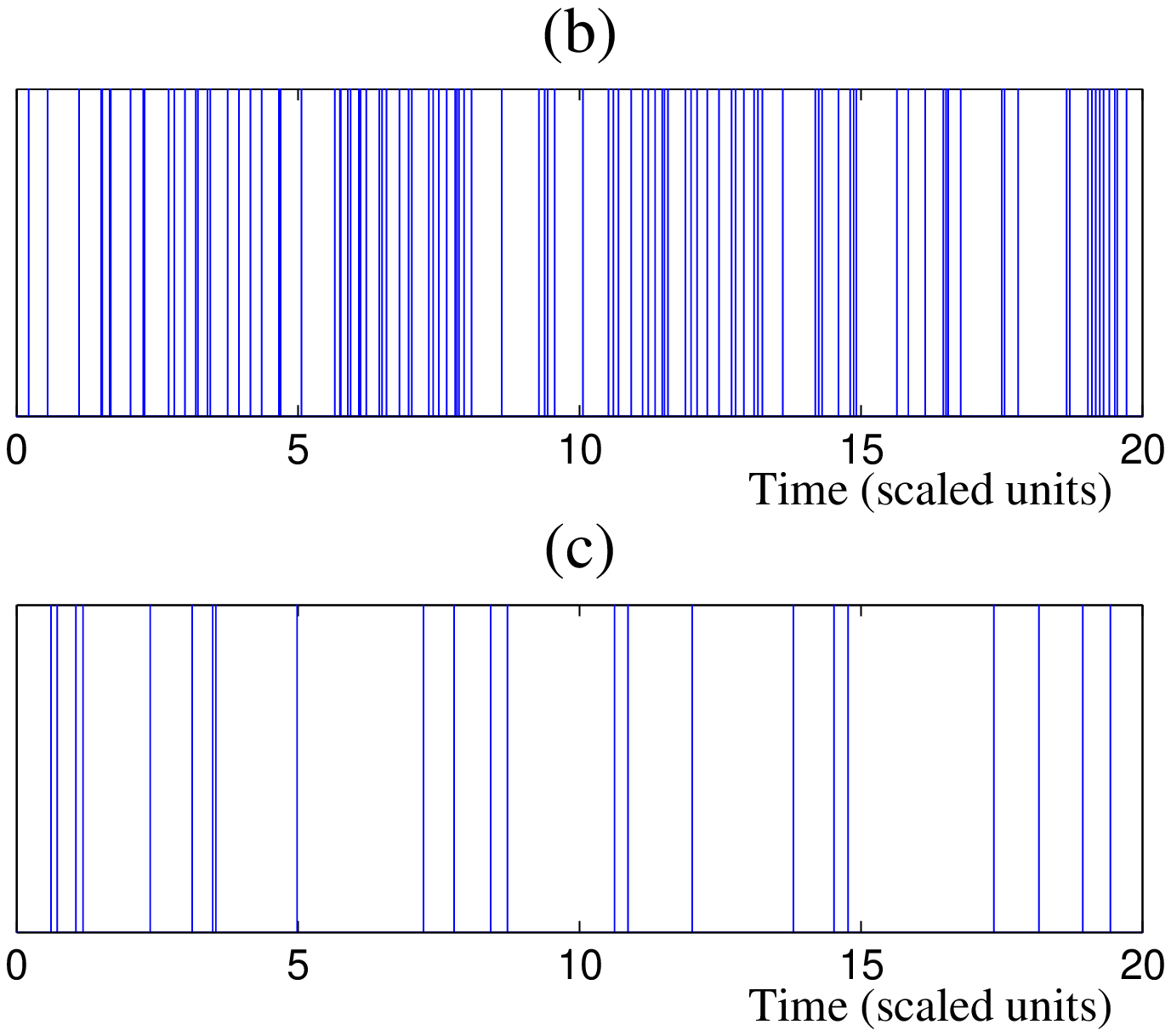}
}
\caption{
(a) Example of a phase space trajectory for $r=1$ with initial state $\ket{\phi^+(0)}=\ket{00}$ and $\epsilon=0.5$;
the system eventually settles into the $\ket{\Phi^-}\leftrightarrow\ket{\Psi^-}$ entangled-state cycle.
(b)~and~(c) Photon counts for Detectors 1 and 2, respectively. 
}\label{fig:fig9}
\end{center}
\end{figure}

\begin{figure}[t]
\begin{center}
\vbox{
\includegraphics[scale=0.4]{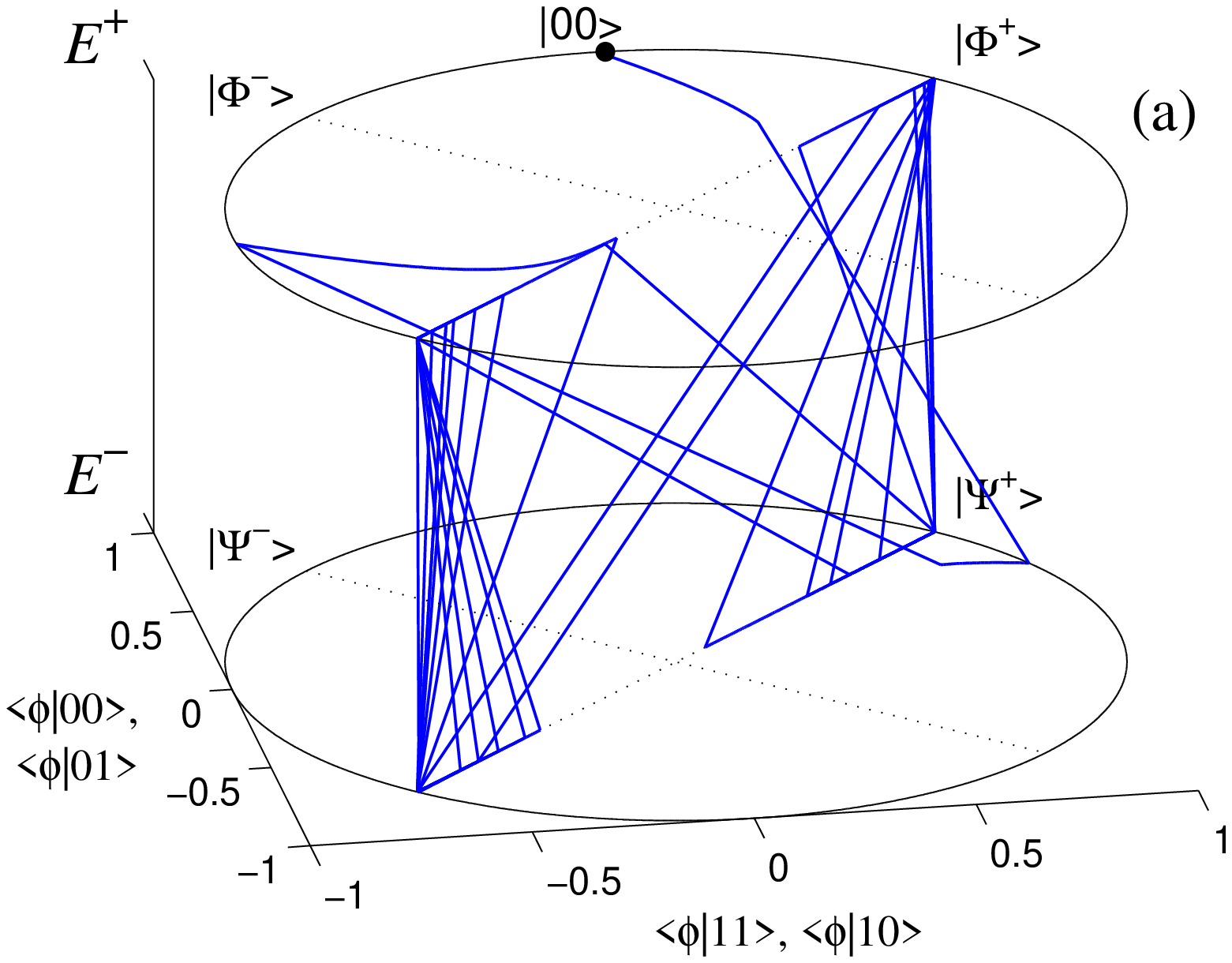}
\vskip0.5cm
\includegraphics[scale=0.4]{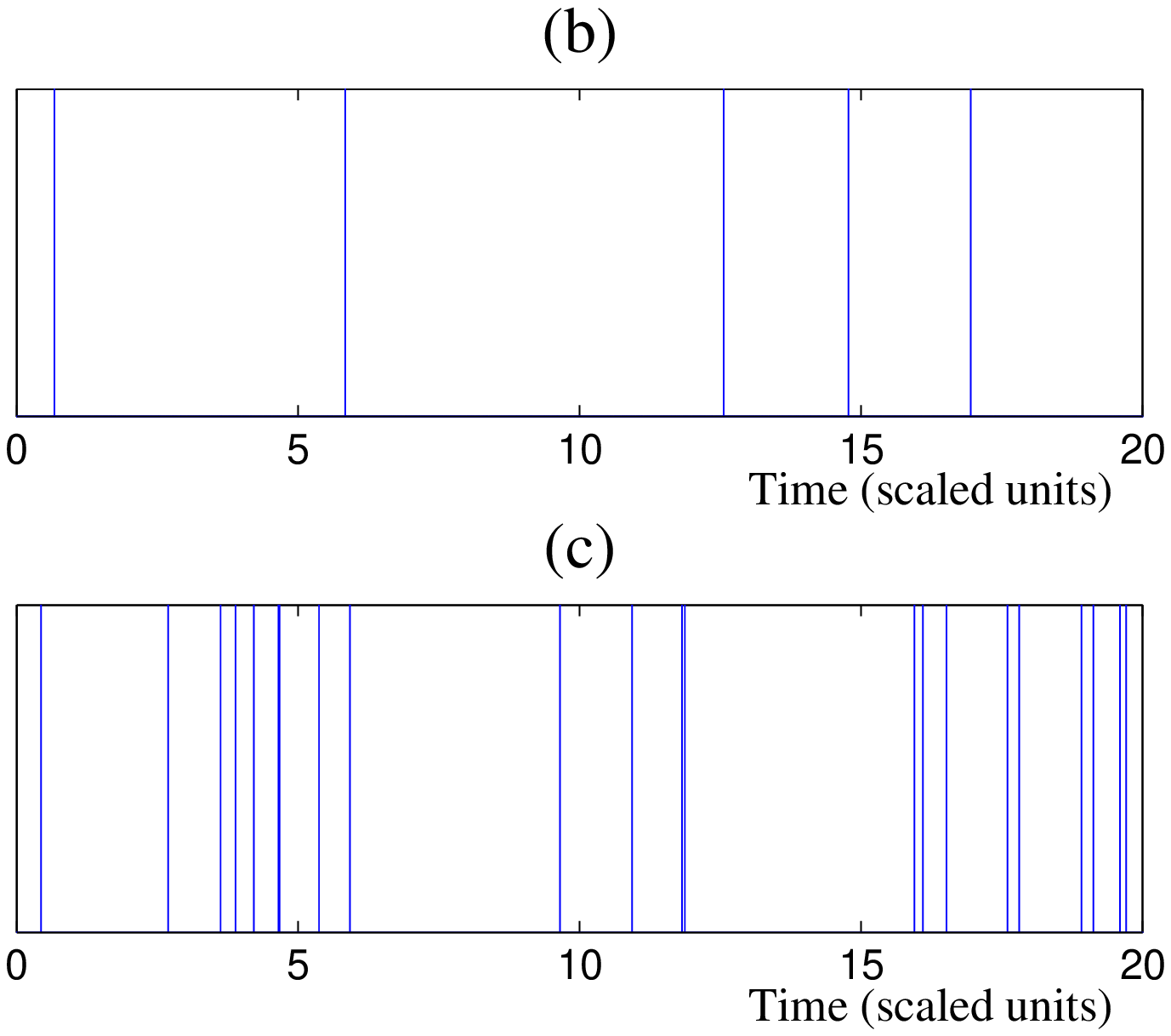}
}
\caption{
(a) Example of a phase space trajectory for $r=1$ with initial state $\ket{\phi^+(0)}=\ket{00}$ and $\epsilon =0.5$; 
the system eventually settles into the $\ket{\Phi^+}\leftrightarrow\ket{\Psi^+}$ entangled-state cycle.
(b)~and~(c) Photon counts for Detector 1 and 2, respectively.
}\label{fig:fig10}
\end{center}
\end{figure}

Our original model allowed for the possibility of imperfect intercavity coupling, through the parameter
$\epsilon$ and the jump operator $\hat C_2$ which describe the effects of photon loss in propagation between
the two cavities. Focusing on the resonant case ($r=1$), we now consider the situation in which $\epsilon <1$.
Typical trajectories for $\epsilon =0.5$ are shown in Figs.~\ref{fig:fig9}(a) and \ref{fig:fig10}(a), with
the two photon count records shown in frames (b) and (c) of the figures. Remarkably, entangled-state cycles
persist, but now the system settles into one or other of two distinct cycles, involving either the symmetric
or antisymmetric Bell states.

To understand the behavior, consider the forms of the operators involved; in particular, for $r=1$, we have
effective Hamiltonian
\begin{equation}
-i\hat H_{\rm eff}= 
\left(\begin{array}{cccc}
-2&0&0&2\sqrt{\epsilon}\\
0&-2&2\sqrt{\epsilon}&0\\
0&2\sqrt{\epsilon}&-2&0\\
2\sqrt{\epsilon}&0&0&-2 \\
\end{array}\right),
\end{equation}
and jump operators
\begin{subequations}
\begin{equation}
\hat C_1= \sqrt{2}\left(\begin{array}{cccc}
0&-1&\sqrt{\epsilon}&0\\
-1&0&0&\sqrt{\epsilon}\\
\sqrt{\epsilon}&0&0&-1\\
0&\sqrt{\epsilon}&-1&0\\
\end{array}\right),
\end{equation}
and
\begin{equation}
\hat C_2=\sqrt{2(1-\epsilon)}\left(\begin{array}{cccc}
0&0&1&0\\
0&0&0&1\\
1&0&0&0\\
0&1&0&0\\
\end{array}\right) .
\end{equation}
\end{subequations}
Significantly, these operators commute with one another,
\begin{eqnarray}
[\hat C_1,\hat C_2]=[\hat C_1,\hat H_{\rm eff}]=[\hat C_2,\hat H_{\rm eff}]=0.
\end{eqnarray}
Their operation upon the Bell states is given by
\begin{subequations}
\begin{eqnarray}
-i\hat H_{\rm eff}\ket{\Phi^\pm}&=&-2\left(1\mp\sqrt{\epsilon}\right)\ket{\Phi^\pm}
=\lambda_\pm\ket{\Phi^\pm},\mkern30mu
\label{Heff1}
\\
-i\hat H_{\rm eff}\ket{\Psi^\pm}&=&-2\left(1\mp\sqrt{\epsilon} \right)\ket{\Psi^\pm}
=\lambda_\pm\ket{\Psi^\pm},
\end{eqnarray}
\end{subequations}
and
\begin{subequations}
\begin{eqnarray}
\hat C_1\ket{\Phi^\pm}&=&(\lambda_\pm/\sqrt2\,)\ket{\Psi^\pm},
\\
\hat C_1\ket{\Psi^\pm}&=&(\lambda_\pm/\sqrt2\,)\ket{\Phi^\pm},
\\
\hat C_2\ket{\Phi^\pm}&=&\pm\sqrt{2(1-\epsilon)}\ket{\Psi^\pm},
\\
\hat C_2\ket{\Psi^\pm}&=&\pm\sqrt{2(1-\epsilon)}\ket{\Phi^\pm}. 
\label{C2Psi}
\end{eqnarray}
\end{subequations}
Thus, the Bell states are eigenstates of $\hat H_{\rm eff}$, and the jump operators interchange Bell states in
$E^+$ and $E^-$: each jump operator converts the symmetric (antisymmetric) Bell state in $E^+$ to the symmetric
(antisymmetric) Bell state in $E^-$ and vice-versa. 

Now, let us consider a particular quantum trajectory for which a total of $n$ jumps occur, separated by the time
intervals $\{\Delta t_i:i=1,\ldots,n\}$. For an initial state $\ket{\phi_0}$, the (unnormalized) state at the
conclusion of the $n$ jumps is written as
\begin{eqnarray}
\ket{\phi_t}=\hat J_n e^{-i\hat H_{\rm eff}\Delta t_n}\ldots\hat J_2 e^{-i\hat H_{\rm eff}
\Delta t_2}\hat J_1 e^{-i\hat H_{\rm eff}\Delta t_1}\ket{\phi_0},\nonumber
\end{eqnarray}
where each $\hat J_i$ is either $\hat C_1$ or $\hat C_2$. Since all operators in the string acting on $\ket{\psi_0}$
commute, this expression can be rewritten in a variety of forms, two of which prove to be especially useful in
explaining the distinct behaviors illustrated by Figs.~\ref{fig:fig9} and \ref{fig:fig10}. In the first case,
we may write
\begin{subequations}
\begin{eqnarray} \label{phit1}
\ket{\phi_t}_{\rm (i)}=\hat C_2^m e^{-i\hat H_{\rm eff}t}\left(\hat C_1^l\ket{\phi_0}\right) ,
\end{eqnarray}
passing all $l$ ocurrences of $\hat C_1$ to the right and all $m$ occurrences of $\hat C_2$ to the left ($l+m=n$);
in the second we write
\begin{eqnarray} \label{phit2}
\ket{\phi_t}_{\rm (ii)}=\hat C_1^l\hat C_2^m \left(e^{-i\hat H_{\rm eff}t}\ket{\phi_0}\right) ,
\end{eqnarray}
where all jump operators are passed to the left.
\end{subequations}

The arbitrary (pure) initial state can be expressed as a superposition of Bell states,
\begin{eqnarray}
\ket{\phi_0}=a\ket{\Phi^+}+b\ket{\Phi^-}+c\ket{\Psi^+}+d\ket{\Psi^-},
\end{eqnarray}
where $a$, $b$, $c$, and $d$ are expansion coefficients, generally complex. Substituting this expansion into
Eqs.~(\ref{phit1}) and (\ref{phit2}), and using Eqs.~(\ref{Heff1})--(\ref{C2Psi})---assuming for simplicity
that $l$ and $m$ are even---the two forms for the state $\ket{\phi_t}$ are
\begin{subequations}
\begin{eqnarray}
\ket{\phi_t}_{\rm (i)}&\propto&e^{-i\hat H_{\rm eff}t}\ket{\phi_0}^\prime,
\label{eq:cycle1}\\
\ket{\phi_t}_{\rm (ii)}&\propto&\hat C_1^l\ket{\phi_t}^\prime,
\label{eq:cycle2}
\end{eqnarray}
\end{subequations}
where
\begin{subequations}
\begin{eqnarray}
\ket{\phi_0}^\prime
&\equiv&\lambda_+^l(a\ket{\Phi^+}+ c\ket{\Psi^+})+\lambda_-^l(b\ket{\Phi^-}+ d\ket{\Psi^-}),
\nonumber\\
\label{eq:cycle1prime}\\
\ket{\phi_t}^\prime
&\equiv&e^{\lambda_+t}(a\ket{\Phi^+}+ c\ket{\Psi^+})+e^{\lambda_-t}(b\ket{\Phi^-}+d\ket{\Psi^-}).\nonumber
\\\label{eq:cycle2prime}
\end{eqnarray}
\end{subequations}
Observe now that the ratio of the eigenvalues satisfies
\begin{equation}
\lambda_+/\lambda_-=(1-\sqrt{\epsilon})/(1+\sqrt{\epsilon})< 1.
\label{eq:ratio}
\end{equation}
It follows that $\ket{\phi_t}_{\rm (i)}$ and $\ket{\phi_t}_{\rm (ii)}$ allow us to predict quite distinct
asymptotic behaviors for the system state. For sufficiently large $l$, the contribution to $\ket{\phi_t}_{\rm (i)}$
from the symmetric Bell states is negligible compared with the contribution from the antisymmetric Bell states,
in which case, using Eqs.~(\ref{eq:cycle1}) and (\ref{eq:cycle1prime}),
\begin{equation}
\ket{\phi_t}_{\rm (i)}\sim e^{\lambda_-t} \lambda_-^l \left(b\ket{\Phi^-}+d\ket{\Psi^-} \right).
\label{eq:asym}
\end{equation}
The system is locked into a cycle between the two antisymmetric Bell states, the situation illustrated in
Fig.~\ref{fig:fig9} (for $\epsilon=0.5$, $\left|\lambda_+/\lambda_-\right|=0.17$). In contrast, for sufficiently
large $t$, the contribution to $\ket{\phi_t}_{\rm (ii)}$ from the antisymmetric Bell states is negligible compared
with that from the symmetric Bell states, and using Eqs.~(\ref{eq:cycle2}) and (\ref{eq:cycle2prime}),
\begin{equation}
\ket{\phi_t}_{\rm (ii)}\sim e^{\lambda_+t}\left(\frac{\lambda_+}{\sqrt{2}}\right)^l
\left(b\ket{\Phi^+}+d\ket{\Psi^+}\right).\label{eq:sym}
\end{equation}
The system is locked into a cycle between the two symmetric Bell states, as shown in Fig.~\ref{fig:fig10}.

Which of the two cycles is chosen in a particular realization of the photon counting record is random, as is
the time taken to settle into the cycle. Effectively, the decision is the outcome of a competition between the
periods of evolution between quantum jumps and the jumps themselves---specifically, those associated with
photon counts at Detector 1. Considering Eqs.~(\ref{eq:cycle1prime}) and (\ref{eq:ratio}), we see that every
count at Detector~1 results in an increased probability to find the system in one of the antisymmetric Bell
states. On the other hand, from Eqs.~(\ref{eq:cycle2prime}) and (\ref{eq:ratio}), the periods of evolution
between counts have the reverse effect---they increase the probability for the system to be found in a
symmetric Bell state. The critical factor that decides which tendency wins is the number of photon counts
occuring at Detector~1 over a given (substantial) interval of time. If there are many, as in Fig.~\ref{fig:fig9}(b),
the entangled-state cycle between antisymmetric Bell states wins out; if there are few, Fig.~\ref{fig:fig10}(b),
the cycle between symmetric Bell states occurs. The same decision mechanism is observed in other examples
\cite{Carmichael94}. Note that counts at Detector 2 are not involved---not directly at least. They do figure
indirectly as a mechanism reducing the average number of counts at Detector 1; indeed, they are the ultimate
source of the asymmetry reflected in the ratio $\lambda_+/\lambda_-<1$.

As the system approaches a particular cycle the quantum trajectory evolution tends to reinforce the establishment
of the cycle. Close to the antisymmetric cycle, the evolution between jumps is dominantly governed by $\lambda_-
=-2(1+\sqrt{\epsilon})$ and is therefore relatively fast. This leads to frequent photon counts at Detector 1
[Fig.~\ref{fig:fig9}(b)]. Close to the symmetric cycle, the between-jump evolution is dominantly governed by
$\lambda_+=-2(1-\sqrt{\epsilon})$, hence is relatively slow. Photon counts at Detector 1 become much less frequent
[Fig.~\ref{fig:fig10}(b)].

From the dramatic difference in count rates at Detector~1 for the two cycles, it is clear that one can determine
which entanglement cycle the system evolves to for a particular realization. However, without knowledge of the
record of photon counts at Detector 2, which by definition we do not have, one can not know where on the cycle
the system is, i.e., whether the state is in $E^+$ or $E^-$. Thus, the ensemble average state of the system is
mixed, described by one of the density operators
\begin{eqnarray}
\rho_\pm=\frac{1}{\sqrt{2}} \left( \ket{\Phi^\pm}\bra{\Phi^\pm} + \ket{\Psi^\pm}\bra{\Psi^\pm} \right).
\end{eqnarray}

\section{Discussion and Conclusions}
\label{subsec5_1}

Consider a thought experiment where the cascaded qubit system, set to resonance, evolves freely and its entire
output is collected and stored inside a black box. At some time the lasers driving the Raman transitions are
turned off, so the evolution ceases. The box and qubits are separated and moved to causally disconnected
regions of space time. Let Alice and Bob be standard observers of the qubits, and give Eve jurisdiction over
the box.

We can now ask, how much entanglement exists between the qubits of Alice and Bob? While this is simply a
roundabout way of asking how entanglement evolves, it helps elucidate some of the key concepts behind the
quantum trajectory measure of entanglement. Conventional entanglement measures are based upon an analysis
of the density matrix at this time. They throw away the box and look at the system of qubits alone---they
disregard Eve and view the system from the perspective of Alice and Bob.

Yet in general every interaction between two objects entangles them, and as the qubit system and box interacted
in the past, their states are intertwined. Neither possess an independent reality, and neither, considered
alone, can be completely described. Eve's box contains information, which, if discarded, adds entropy to the
qubit system of Alice and Bob. This entropy is the source of ambiguity in the quantification of entanglement.
From this point of view, as noted in the introduction, the problem of bi-partite entanglement in an open system
relates to that of tri-partite entanglement in a closed one. To completely characterize the entanglement of the
present example, in addition to the entanglement between Alice and Bob, we must consider their entanglement with
Eve.

A quantum description of the box is impractical, but it is feasible to extract classical information about
what it contains, through measurement. Quantum trajectories facilitate this, and allow us not to discard the
box completely. In turn, the system state retains its purity, conditional on the classical information
extracted from the box. With this extra information, we can extract more entanglement from the cascaded qubit
system.

Working from the master equation for the cascaded system \cite{Clark03}, previously it was assumed that the
system evolved gradually into a pure state, whereby entanglement was generated. The behaviour at resonance,
however, was unclear, since there the master equation had two zero eigenvalues and no well-defined steady
state. By considering the conditional evolution we have shown that, at resonance, asymptotically the system
is either in the Bell state $\ket{\Phi^+}$ or oscillating (stochastically switching) between two Bell states,
$\ket{\Phi^-}$ and $\ket{\Psi^-}$.

From the density matrix point of view, the latter is an equal mixture of Bell States and would yield no
entanglement under any mixed state measure; physically, Alice and Bob, without collaboration from Eve,
cannot extract any entanglement from their qubits. Suppose, however, that Eve opens her box to count the
number of photons inside. Seeing whether the count is even or odd, she is able to deduce exactly which
Bell state Alice and Bob's system is in. Thus, her measurement unravels the density operator, creating
entanglement, despite the fact that the measurement is not causally connected to Alice and Bob's qubits.

It is tempting to say that the entanglement was always there, as a matter of fact, until one realizes that
there are many other ways in which Eve could choose to measure her state, each producing a different
unravelling of the qubit system and yielding a different value of entanglement. The entanglement facilitated
by Eve's measurements is \textit{contextual} in this sense.

This thought experiment demonstrates why any attempt to quantify the entanglement of an open system from
the density operator alone cannot be considered complete. The density operator should not be treated as a
fundamental object, as it does not provide a complete description of the physical state. We have presented
a simple example where oscillations between maximally entangled states are hidden within a separable
density operator. The fact that the density operator contains entropy, implies that information about
its entanglement with an external system was discarded at some time. In studying such a mixed state,
there is benefit from considering, not only the mixed state itself, but the process through which it was
generated, and the access this potentially gives to a conditional dynamics.

The results of this paper could be extended by employing quantum trajectories in a broader sense. In cases where
the results of environmental interactions cannot be measured, such as coupling loss, Wiseman and Vaccaro \cite{wiseman1}
have shown that only certain unravelings can be physically realized. A conceivable measure of entanglement would take
the minimum of all physically realizable unravelings. Alternatively, one might take the maximum of all physically
realizable unravellings, which would measure the maximum distillable entanglement when local measurements on
the environment are taken into account. 

This work was supported by the Marsden Fund of the RSNZ.

\end{document}